%% file: arxiv2.tex
\documentclass[11pt,a4paper]{article}

\include{preambleSS}

\title{Four-point functions of all-different-weight chiral primary operators in the supergravity approximation }

\author[a,1]{Gleb Arutyunov}
\note{Correspondent fellow at Steklov
Mathematical Institute, Moscow.}
\author[a]{Rob Klabbers}
\author[a]{and Sergei Savin}
\affiliation[a]{II. Institut f\"ur Theoretische Physik, Universit\"at Hamburg, Luruper Chaussee 149, 22761 Hamburg, Germany\\
Zentrum f\"ur Mathematische Physik, Universit\"at Hamburg, Bundesstrasse 55, 20146 Hamburg, Germany
}
\emailAdd{gleb.arutyunov@desy.de}  
\emailAdd {rob.klabbers@desy.de}
\emailAdd{sergei.savin@desy.de}

\abstract{Recently a Mellin-space formula was conjectured for the form of correlation functions of $1/2$ BPS operators in planar $\mN=4$ SYM in the strong 't Hooft coupling limit. In this work we report on the computation of two previously unknown four-point functions of operators with weights $\langle 2345 \rangle$ and $\langle 3456\rangle$, from the effective type-IIB supergravity action using AdS/CFT. These correlators are novel: they are the first correlators with all different weights and in particular $\langle 3456\rangle$ is the first next-next-next-to-extremal correlator to ever have been computed. We also present simplifications of the known algorithm, without which these computations could not have been executed. These simplifications consist of a direct formula for the exchange part and for the contact part of the correlation function, as well as a simplification of the $C$ tensor algorithm to compute $a$ tensors. After bringing our results in the appropriate form we successfully corroborate the recently conjectured formula. 

}

\begin{document}
\begin{flushright}
\scriptsize{ZMP-HH/18-14}
\end{flushright}

\maketitle
\flushbottom

\renewcommand{\thefootnote}{\arabic{footnote}}
\setcounter{footnote}{0}

\section{Introduction}

Understanding the AdS/CFT correspondence has been one of the major goals in theoretical high-energy physics for almost twenty years \cite{Maldacena:1997re}. One way of pursuing this is the computation of observables on either side of this correspondence. For the well-studied holographic pair made up by $\mN=4$ super Yang-Mills theory (SYM) and type IIB superstring theory on the \ads background these computations are most feasible in the planar regime \cite{tHooft:1973alw, Gubser:1998bc,Witten:1998qj}. A further simplification of the string theory occurs when considering the 't Hooft limit yielding a classical supergravity theory dual to SYM at strong coupling. 

Recently the problem of finding holographic four-point correlation functions in this particular limit has gained renewed interest. In \cite{Rastelli:2016nze, Rastelli:2017udc} the Mellin space formalism was used to analyze the structure of correlation functions of four $1/2$ BPS local operators of arbitrary weights. This culminated in a conjectured closed formula for these correlators in Mellin space, thereby extending the existing formula for equal-weight operators proposed in \cite{Dolan:2006ec}. The conjecture follows from reformulating the computation of four-point functions as a bootstrap problem, imposing certain properties such as superconformal symmetry on the correlators. One property, namely linear asymptotic growth of the correlator at large values of the Mandelstam variables, was proven shortly after in \cite{Arutyunov:2017dti}. 

The existence of the simple Mellin-space formula is surprising when one considers the only method known at present to explicitly compute these correlation functions: one considers the tree-level Witten diagrams for the chosen operators, whose vertices follow from the effective action of the Kaluza-Klein reduction of the type IIB supergravity on $S^5$ \cite{Arutyunov:1998hf, Arutyunov:1999en, Arutyunov:1999fb}. This computation becomes cumbersome quickly and yields very unwieldy intermediate results, although the final result can usually be expressed in a compact form, in turn suggesting a simple formula such as the one conjectured in \cite{Rastelli:2016nze} might exist. However, because of the sheer difficulty of their computation it is equally possible that we have simply not explored far enough to discover four-point functions whose final form takes a less appealing form. Indeed, as of yet all the explicitly computed four-point functions are in some form not completely generic: the first computed correlators were the equal-weight ones for $1/2$ BPS operators with (lowest) weight $k=2$ \cite{DHoker:1999kzh, Arutyunov:2000py}, $k=3$ \cite{Arutyunov:2002fh} and $k=4$ \cite{Arutyunov:2003ae}, where the computation gets simplified due to the internal symmetry. After these correlation functions the first four-point function featuring different-weight operators was discussed in \cite{Berdichevsky:2007xd}, which considers the correlator of two $k=2$ and two $k=3$ operators. This was shortly afterwards generalized to the case in which the $k=3$ operators become arbitrary (but equal) weights \cite{Uruchurtu:2008kp}. The maximal amount of different operators studied until present was three in \cite{Uruchurtu:2011wh}, where next-next-to-extremal correlators were studied that generalize those in \cite{Uruchurtu:2008kp}. Finally, the only other explicitly known result is the equal-weight correlation function of weight $k=5$, which was computed using a bootstrap approach in \cite{Rastelli:2017udc}. These results, although restricted to $1/2$ BPS operators, do in some cases allow for the evaluation of correlation functions of other members of these $1/2$ BPS multiplets. For example, for operators belonging to the stress-energy multiplets it was shown in \cite{Korchemsky:2015ssa} that all their four-point functions can be found from the equal-weight correlator for $1/2$ BPS operators of weight $2$. 

Using the explicit results as a stepping stone it has proven possible to probe the non-planar spectrum of $\mN=4$ SYM, by considering perturbations in $1/N$ around the supergravity results \cite{Alday:2017xua,Aprile:2017xsp,Aprile:2017qoy,Aprile:2018efk}. The newly-occurring mixing problem of double-trace operators can be solved using the results known from pure supergravity. Interestingly, the results show the existence of a degeneracy of the supergravity spectrum that gets lifted once one considers loop corrections. On the other hand these results provide an insight into the full spectrum of $\mN=4 $ SYM, which is still beyond our reach.  

In fact, even the known supergravity correlators are degenerate in one form or another: they are either particularly symmetric (equal-weights) or close to extremality and sometimes both. In fact, not a single non-trivial correlation function has ever been computed between four operators with four different weights. Also, apart from the equal weight case for which an explicit formula has long ago been conjectured \cite{Dolan:2006ec}, no four-point functions which go beyond the next-next-to-extremal case have been considered. In this paper we do precisely these two things by presenting the computation of the $\langle 2345 \rangle$ and $\langle 3456 \rangle$ correlators, where each digit represents the weight of an operator, using the effective supergravity action. These four-point functions consist of all-different operators and $\langle 3456 \rangle$ is also next-next-next-to-extremal, therefore no longer enjoying simplifications due to being close to extremality. 

A particular -- although not completely new-- feature of these correlation functions is the effect of the presence of extended operators: it has been known for a long time that in general the supergravity scalar fields are not dual to the single-trace operators from the $1/2$ BPS multiplets, but rather to a linear combination of single- and double-trace operators known as extended operators \cite{Arutyunov:2000ima}. Nevertheless, for most purposes this fact can be ignored as in the supergravity limit it only affects the free part of the correlators containing at least weight-$4$ operators. In fact, the only known correlators for which this effect was seen are those from the family $\langle 22 nn \rangle$ for $n>3$ discussed in \cite{Uruchurtu:2008kp}.

Our computations serve multiple goals: first of all we present simplifications to the algorithm that computes four-point correlation functions from the supergravity action that, given sufficient computing power, should allow for the computation of further non-trivial correlators. This algorithm -- being a refined version of the procedure that has been used to compute all the correlators \cite{DHoker:1999kzh, Arutyunov:2000py,Arutyunov:2002fh,Arutyunov:2003ae,Berdichevsky:2007xd,
Uruchurtu:2008kp, Uruchurtu:2011wh} -- is fully rigorous and does not include any bootstrapping. In particular, the simplifications include an explicit formula for the exchange part in terms of Witten diagrams and a drastic simplification of the computation of the vertex building blocks. These might even serve as a start to prove the aforementioned Mellin-space conjecture. Secondly, we use these simplifications to compute two explicit correlators and use them to check the Mellin-space conjecture from \cite{Rastelli:2016nze, Rastelli:2017udc} further away from symmetry and extremality. We indeed find agreement with the conjecture. 

This paper is set up as follows: in section \ref{sec:generalities} we recall the general structure of four-point functions of $1/2$ BPS operators in $\mN=4$ SYM, introducing much of the needed machinery. In section \ref{sec:supergravity} we discuss the known method how to compute these four-point functions from the effective supergravity action and present simplifications, in particular to the exchange terms. In section \ref{sec:results} we present our main result, the explicit form of the $\langle 2345 \rangle$ and $\langle 3456 \rangle$ correlators. Subsequently, in section \ref{sec:analyzing}, we discuss how we check the correctness of the correlators and we show that our results are consistent with the Mellin conjecture from \cite{Rastelli:2017udc} and can be further simplified using knowledge of their Mellin-space form. This is followed by a detailed discussion in section \ref{sec:extendedness} of extended operators and the free parts of their correlation functions, including a proof for which cases this effect is relevant. In section \ref{sec:discussion} we conclude and discuss future directions. 

We gather many of the details of the computation in the appendix: in appendix \ref{app:C-algebra} we discuss how to compute the vertices in the supergravity Lagrangian using C-algebra. In particular we discuss how one can use the internal symmetries of some of the vertex building blocks (known as $a$ tensors) to drastically simplify their computation by introducing a decomposition of the symmetric group. We furthermore present new relations between the $a$'s and the other building blocks ($t$ and $p$ tensors) in the form of reduction formulae that allow for a further simplification of the algorithm. Finally, in appendix \ref{app:intermediate results} we gather many of the intermediate results that are of general interest. These include the computation of explicit SO$(6)$ Clebsch-Gordan coefficients, intermediate forms of the $\langle 2345 \rangle$ and $\langle 3456 \rangle$ as they follow from our studies and planar free correlators for all non-trivial correlators containing at most weight $5$ operators.

\section{General structure of four-point function of CPOs}\label{sec:generalities}
The central objects of study in this paper are the four-point correlation functions of $1/2$ BPS operators in planar $\mN=4$ SYM theory at large values of the 't Hooft coupling constant $\lambda$. We will start with a review of the structure of these correlation functions -- due to the presence of superconformal symmetry -- here, see e.g. \cite{Chicherin:2015edu}, as it will introduce a lot of our notation. 
\\[5mm]
The multiplets of $1/2$ BPS operators are generated from their highest weight vectors called conformal primary operators (CPO) which we will denote as $\tilde{\mO}_k$ with $k$ indicating the conformal dimension of the CPO. They can be decomposed as
\be
\label{eq:BPSdecomp}
\tilde{\mO}_k = \tilde{\mO}^{i_1\ldots i_k} t_{i_1} \cdots t_{i_k},
\ee
with $ \tilde{\mO}^{i_1\ldots i_k}$ carrying the dependence on the scalar fields $\phi^i$ and $t$ being a six-dimensional null vector that keeps track of the SO$(6)$ $R$-symmetry of the operator. Given four weights $(k_1,k_2,k_3,k_4)$ with $k_i \geq 2$ the four-point correlation function of such operators is given by
\be
\label{eq:correlator}
\langle k_1 k_2 k_3 k_4 \rangle \coloneqq \langle \tilde{\mathcal{O}}_{k_1}\left(x_1,t_1 \right)\cdots\tilde{\mathcal{O}}_{k_4}\left(x_4,t_4 \right) \rangle,
\ee
where the $x_i$ are spacetime coordinates and we indicated the dependence on the null vectors $t_1,\ldots , t_4$.\footnote{Note that the subscript does not refer to a vector component, but is there to distinguish four different $t$ vectors.} The correlator splits into a free part and an interacting part:
\begin{equation}
\label{eq:formofthecorrelator}
\langle k_1 k_2 k_3 k_4 \rangle = \langle k_1 k_2 k_3 k_4 \rangle_{0} + \langle k_1 k_2 k_3 k_4 \rangle_{\text{int}},
\end{equation}
where the first part $\langle k_1 k_2 k_3 k_4 \rangle_{0}$ is the Born approximation and can be computed by considering Wick contractions. After defining 
\begin{equation}
d_{ij} = \frac{t_{ij}}{x_{ij}^2}, \quad \text{ with } t_{ij} = t_i \cdot t_j,\quad x_{ij}^2 = (x_i-x_j)^2
\end{equation}
we can state the form of the free part as:
\begin{equation}
\label{eq:freeG}
\langle k_1 k_2 k_3 k_4 \rangle_{0} = \sum_{a}c_{a} \left( \prod_{i,j} d_{ij}^{a_{ij}} \right),
\end{equation}
where the sum runs over all partitions $a$ where the $c_a$ are constants. The partitions parametrize all the inequivalent ways one can pair the $t$ vectors and can be represented as symmetric $4\times4$ matrices with non-negative integer entries with zeroes on the diagonal that satisfy for each $i$
\begin{equation}
\label{eq:aequations}
\sum_{\substack{j=1 \\ j\neq i}}^4 a_{ij} = k_i. 
\end{equation}
For a given set of weights these equations have a finite number of solutions. For later convenience we define 
\begin{equation}
\label{eq:Tl}
T_l \equiv \prod_{\substack{i,j=1 \\ i<j}}^4 t_{ij}^{\left(a_l\right)_{ij}},
\end{equation}
where $a_l$ is the $l$th solution to the equations \eqref{eq:aequations} in the ordered list of solutions: we order solutions $\{a_{12},a_{13},a_{14},a_{23},a_{24},a_{34}\}$ by a lexicographical ordering of these numbers with the lowest one occurring first. In fact, the entire correlator can be written in the form \eqref{eq:freeG}, where the constants $c_a$ then become functions of the conformal cross ratios
\begin{equation}
u= \frac{x_{12}^2x_{34}^2}{x_{13}^2x_{24}^2},\quad v= \frac{x_{14}^2x_{23}^2}{x_{13}^2x_{24}^2}.
\end{equation}
As it turns out, due to supersymmetry the correlator factorizes further: we can write the interacting part as
\begin{equation}
\label{eq:interacting}
\langle k_1 k_2 k_3 k_4 \rangle_{\text{int}} =  \frac{\mathcal{R}_{1,2,3,4}}{x_{13}^2x_{24}^2} \sum_{b} \left( \prod_{i,j} d_{ij}^{b_{ij}} \right)\mF_{b}(u,v),
\end{equation}
where 
\begin{equation}
\label{eq:Rpolynomial}
\begin{aligned}
\mathcal{R}_{1,2,3,4} &= d_{12}^2d_{34}^2x_{12}^2x_{34}^2 + d_{13}^2d_{24}^2x_{13}^2x_{24}^2 + d_{14}^2d_{23}^2x_{14}^2x_{23}^2 \\
&+ d_{12}d_{23}d_{34}d_{14}\left(x_{13}^2 x_{24}^2 - x_{12}^2x_{34}^2 -x_{14}^2 x_{23}^2    \right) \\ 
&+ d_{12}d_{13}d_{24}d_{34}\left(x_{14}^2 x_{23}^2 - x_{12}^2x_{34}^2 -x_{13}^2 x_{24}^2    \right) \\
&+ d_{13}d_{14}d_{23}d_{24}\left(x_{12}^2 x_{34}^2 - x_{14}^2x_{23}^2 -x_{13}^2 x_{24}^2    \right) 
\end{aligned}
\end{equation}
is the fully symmetric general prefactor \cite{Eden:2000bk}. This time the partitions $b$ form a symmetric $4\times4$-matrix with zeroes on the diagonal, but they satisfy the modified equations
\begin{equation}
\label{eq:bequations}
\sum_{\substack{j=1 \\ j\neq i}}^4 b_{ij} = k_i-2. 
\end{equation}
This reduces the number of partitions $b$ significantly, implying that the correlator can be written as a sum over only a few independent functions $\mF_b$ of $u$ and $v$, which are moreover conformally invariant by construction and carry all the dependence on the 't Hooft coupling. 

We will use this decomposition of the correlator for two purposes: firstly to be able to present our computed correlators in a compact form $\{\mF_b\}$. Secondly, this provides a check whether these correlators are consistent with superconformal symmetry. 
\paragraph{$C$ tensors.} The $\mathcal{N}=4$ SYM four-point functions can be represented in many forms depending on the context. In this paper we will use the language of $C$ tensors to compute some of the necessary intermediate objects\footnote{It is because of the necessary appearance of $C$ tensors that we do not follow the notation used in e.g. \cite{Rastelli:2016nze, Rastelli:2017udc} that would allow us to write correlators as polynomials in "$R$ symmetry cross ratios" $\sigma$ and $\tau$: we feel this would lead to an even bigger proclivity of notations.}. The $C$ tensors can be used to track the SO$(6)$ symmetry instead of the $t$ vectors, by writing \eqref{eq:BPSdecomp} as
\be
\tilde{\mO}_k = \tilde{\mO}^{i_1\ldots i_k} C_{i_1,\ldots,i_k},
\ee
with the tensor $C$ being totally symmetric and traceless in its indices. The vector space of $C$ tensors with $k$ indices forms a representation space for the traceless symmetric SO$(6)$ representation with Dynkin labels $[0,k,0]$. Choosing a basis in this space amounts to choosing a set of $C$ tensors which we label with an upper index $I_k$ that will appear throughout this paper:
\begin{equation}
C_{i_1,\ldots,i_k}^{I_k}, \quad \text{with } I_k = 1, \ldots , \text{dim}([0,k,0]). 
\end{equation}
The possible contractions of four $C$ tensors are characterized by the number of connections between the different $C$ tensors. Denoting the number of connections by $a_{ij}$ they should satisfy $a_{ij}=a_{ji}$, $a_{ii}=0$ and the sum conditions for $a$'s \eqref{eq:aequations}, showing that the $a$ partitions indeed parametrize the possible tensor structures \eqref{eq:Tl} for a given correlator \eqref{eq:correlator}. 
%
\section{$\mN=4$ SYM four-point functions from type IIB supergravity}
\label{sec:supergravity}
A method to compute the correlation functions from the previous section using the AdS/CFT correspondence has been known for a long time \cite{Witten:1998qj, Gubser:1998bc}: chiral primary operators of 4D \N SYM theory are dual to the Kaluza-Klein modes of type IIB supergravity on \ads after compactification on $S^5$, which are usually denoted as $s^{I_k}_k$, where as before the index $I_k$ runs over the basis of the corresponding SO$(6)$ representation with Dynkin labels $[0,k,0]$. One computes the on-shell string partition function $\exp(-S_{IIB})$ as a function of the boundary values of the fields $s^{I_k}_k(\vec{x})$ and subsequently interprets it as a generating functional for correlation functions of CPOs whose sources are $s^{I_k}_k(\vec{x})$. Thus, in practice the correlation functions can be determined from the expression:
\bea
\label{eq:corrWitten}
 \langle \tilde{O}^{I_1}_{k_1}(\vec{x}_1) \tilde{O}^{I_2}_{k_2}(\vec{x}_2) \tilde{O}^{I_3}_{k_3}(\vec{x}_3) \tilde{O}^{I_4}_{k_4}(\vec{x}_4) \rangle = \left. \frac{\de^4}{\de s^{I_1}_{k_1}(\vec{x}_1) \de s^{I_2}_{k_2}(\vec{x}_2) \de s^{I_3}_{k_3}(\vec{x}_3) \de s^{I_4}_{k_4}(\vec{x}_4)} \exp(-S_{IIB}) \right|_{s_k\to 0}.
 \nonumber\\
\eea
To compute an $n$-point function it is necessary to expand the action up to the $n$-th order in perturbations of the fields around their background values. For the four-point case explicit expressions were derived in \cite{Lee:1998bxa,Arutyunov:1998hf,Arutyunov:1999en,Arutyunov:1999fb} and the relevant part of the action may be written in the form:
\bea
\label{eq:adsaction}
	S=\frac{N^2}{8\pi^2}\int[dz](\mathcal{L}_2+\mathcal{L}_3+\mathcal{L}_4).
\eea
Here we use the Euclidean \AdS\, metric $[dz]=\dfrac{d^5z}{z_0^5}$, which leads to a change of the overall sign for the action. The langrangian splits into a quadratic part $\mathcal{L}_2$, a cubic part $\mathcal{L}_3$ and a quartic part $\mathcal{L}_4$. The first two contribute to the so-called \emph{exchange part} of the correlator, because they give rise to exchange interactions captured in Witten diagrams such as in fig. \ref{fig:Witten}, while $\mathcal{L}_4$ forms the \emph{contact part} as they lead to contact interactions (see fig. \ref{fig:Wittencontact}). We will discuss how one can compute these two parts for a concrete set of weights separately, beginning with the latter.  
%
\subsection{Contact part}
\label{sec:Contact}
The quartic Lagrangian is given by:
\bea
\label{eq:4quartic}
	&&\mathcal{L}_4=\mathcal{L}_4^{(4)}+\mathcal{L}_4^{(2)}+\mathcal{L}_4^{(0)}, \nonumber\\
	&&\mathcal{L}_4^{(4)} = (S^{(4)}_{I_1 I_2 I_3 I_4}+A^{(4)}_{I_1 I_2 I_3 I_4})~s^{I_1}\na_\m s^{I_2} \na_\nu^2 (s^{I_3}\na^\m s^{I_4}),\nonumber\\
	&&\mathcal{L}_4^{(2)} = (S^{(2)}_{I_1 I_2 I_3 I_4}+A^{(2)}_{I_1 I_2 I_3 I_4})~s^{I_1}\na_\m s^{I_2} s^{I_3}\na^\m s^{I_4}, \\
	&&\mathcal{L}_4^{(0)} = (S^{(0)}_{I_1 I_2 I_3 I_4}+A^{(0)}_{I_1 I_2 I_3 I_4})~s^{I_1}s^{I_2}s^{I_3}s^{I_4}. \nonumber
\eea
The repeated indices $I_1,\ldots , I_4$ here imply the summation over all possible $[0,k,0]$ representations as well as summation over the basis of the representation space: 
\begin{equation}
\label{eq:summationconvention}
g_I s^I\equiv \sum\limits_{k\geq 2}\sum\limits_{I_k=1}^{\text{dim}([0,k,0])}g_{I_k}s_k^{I_k}.
\end{equation}
The various couplings $S$ and $A$ \cite{Arutyunov:1999fb} depend on contractions of the previously discussed $C$ tensors, which are Clebsch-Gordan coefficients for the tensor product of SO$(6)$ irreps. Their explicit expressions depend on the chosen weights and their computation becomes complicated quite quickly. In fact, this forms the main difficulty in the computation of higher-weight correlators in this formalism. We have found ways to simplify this computation considerably, but their discussion is fairly technical. We therefore refer the reader to the discussion in appendix~\ref{app:C-algebra}.

Further simplifications follow from a closer analysis of the four-derivative terms $\mathcal{L}_4^{(4)}$: it was shown in \cite{Arutyunov:2000ima} that these terms can be made to vanish in the extremal $k_1=k_2+k_3+k_4$ and sub-extremal $k_1=k_2+k_3+k_4-2$ cases. Moreover, in \cite{Arutyunov:2017dti} the vanishing of the four-derivative terms was proven for all four-point correlators of $1/2$ BPS operators. To show this, one has to perform integration by parts in the expression for $\mathcal{L}_4^{(4)}$, which produces contributions to lower-derivative terms $\mathcal{L}_4^{(4\rightarrow 2)}$ and $\mathcal{L}_4^{(4\rightarrow 0)}$. However, in the same manner as in \citep{Arutyunov:2017dti}, using the reduction formulae  \eqref{eq:redt} and \eqref{eq:redt2}, one can show that the contribution $\mathcal{L}_4^{(4\rightarrow 0)}$ vanishes identically. Thus, the full contact Lagrangian can be written as 
\be 
\label{eq:fullcontact}
\mathcal{L}_4^{(4\rightarrow 2)}+\mathcal{L}_4^{(2)} + \mathcal{L}_4^{(0)},
\end{equation}
where 
\bal
	\mathcal{L}_4^{(4\rightarrow 2)}&=&\left(S_{I_1I_2I_3I_4}^{(4)}\,(m_1^2+m_2^2+m_3^2+m_4^2-4)-4A_{I_1I_2I_3I_4}^{(4)}\right)~s^{I_1}\na_\m s^{I_2} s^{I_3}\na^\m s^{I_4}, 
\eal
where $m_i$ denotes the \AdS~mass of the corresponding scalar field. 

Now one has to compute the contact part of the on-shell action. For this we solve the equations of motion for scalar fields perturbatively: $s^{I_k}_k=\bar{s}^{I_k}_k+\tilde{s}^{I_k}_k$, where $\bar{s}^{I_k}_k$ is the solution of the linearized equations of motion with fixed boundary conditions and corrections $\tilde{s}^{I_k}_k$ correspond to scalars with vanishing boundary conditions. The solution of the boundary problem is given in \cite{Witten:1998qj}:
\bea
\label{eq:boundsol}
	\bar{s}^{I_k}_k(z)=C_k\int d^4\vec{x}\, K_k(z,\vec{x})s^{I_k}_k(\vec{x}),
\eea
where 
the bulk-to-boundary propagator reads as
\bea
	K_k(z,\vec{x}) =  \left({z_0 \ov z_0^2+(\vec{z}-\vec{x})^2}\right)^k.
\eea
For now we will neglect the normalization factors $C_k$ and take them into account when computing the full correlator. According to \cite{Freedman:1998tz}, they are
\bea
\label{eq:coeffCk}
	C_2={1\ov 2\pi^2}~~\text{and}~~C_k={\G(k)\ov \pi^2 \G(k-2)}~~\text{for}~k > 2.
\eea	

\begin{figure}[t]
\centering
\includegraphics[scale=0.7]{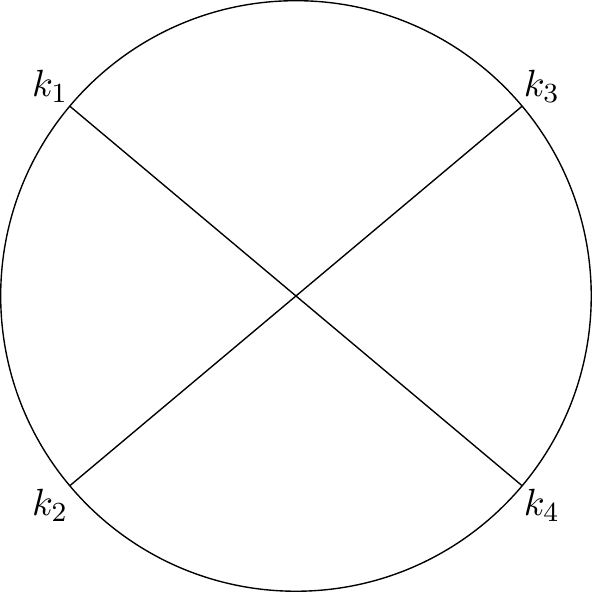}
\caption{A contact Witten diagram.}
\label{fig:Wittencontact}
\end{figure}

To proceed, we define the so-called $D$ functions\footnote{Note, that the set of indices $\{k_1,k_2,k_3,k_4\}$ in the notation for the $D$-functions is always ordered in such a way that it corresponds to an ordered set of variables $\{\vec{x}_1,\vec{x}_2,\vec{x}_3,\vec{x}_4\}$ }, or contact Witten diagrams (see fig. \ref{fig:Wittencontact}), as integrals over $\AdS_5$ space, for details see e.g. \cite{DHoker:1999kzh,Uruchurtu:2008kp}:
\bea
\label{eq:Dfunctions}
	D_{k_1 k_2 k_3k_4} &\equiv & D_{k_1 k_2 k_3k_4}(\vec{x}_1,\vec{x}_2,\vec{x}_3,\vec{x}_4)\nonumber\\
	&=&\int[dz]\,K_{k_1}(z,\vec{x}_1)K_{k_2}(z,\vec{x}_2)K_{k_3}(z,\vec{x}_3)K_{k_4}(z,\vec{x}_4).
\eea
and use the identity 
\bea
	&\na_\m K_{k_1}(z,\vec{x}_1) &\na^\m K_{k_2}(z,\vec{x}_2)=\nonumber\\
	&&k_1k_2 \left(K_{k_1}(z,\vec{x}_1)K_{k_2}(z,\vec{x}_2)-2 |\vec{x}_{12}|^2 K_{k_1+1}(z,\vec{x}_1)K_{k_2+1}(z,\vec{x}_2)\right).\quad 
\eea
To compute a correlator with specific weights one needs to compute the quartic couplings corresponding to all the 24 permutations of the weights. Each of them is multiplied by the correspondingly permuted set of $D$ functions, as follows from carefully going through the steps: to compute a correlator with fixed weights $\{k_1,k_2,k_3,k_4\}$ one restricts the infinite sum in \eqref{eq:4quartic} to representations which correspond to these weights. The sums over $k$'s are not ordered, therefore there are 24 nonzero summands corresponding to the 24 permutations of the indices $\{k_1,k_2,k_3,k_4\}$. In fact, even when some of the weights are equal, potential overcounting is compensated by functional differentiation when computing the correlator. Let us illustrate this procedure on the two-derivative term\footnote{Here for simplicity we write the summation indices $I_1,...,I_4$ as $1,...,4$. The range of these summation indices depend on the weight $k_i$. } $g_{1234}\,s^1s^2\na_\m s^3\na^\m s^4$, where $g_{1234}$ schematically denotes the corresponding coupling:
\bea
&&\int[dz]\,g_{1234}\,s^1s^2\na_\m s^3\na^\m s^4 = \\
	&&\int[dz]\left( g_{1234}\,s_{k_1}^1 s_{k_2}^2 \na_\m s_{k_3}^3 \na^\m s_{k_4}^4 + g_{1234}\,s_{k_3}^1 s_{k_2}^2 \na_\m s_{k_1}^3 \na^\m s_{k_4}^4+...\right)=\int d^4\vec{y}_1 d^4\vec{y}_2 d^4\vec{y}_3 d^4\vec{y}_4\nonumber\\
	&&~~\times \left[ g_{1234}\,k_3k_4\left( D_{k_1 k_2 k_3 k_4}-2|\vec{y}_{34}|^2D_{k_1 k_2 k_3+1 k_4+1}\right)s_{k_1}^1(\vec{y}_1) s_{k_2}^2(\vec{y}_2)s_{k_3}^3(\vec{y}_3) s_{k_4}^4(\vec{y}_4)\right.\nonumber\\ 
	&&~~~+ \left. g_{1234}\,k_1k_4\left( D_{k_3 k_2 k_1 k_4}-2|\vec{y}_{14}|^2D_{k_3 k_2 k_1+1 k_4+1}\right)s_{k_3}^1(\vec{y}_1) s_{k_2}^2(\vec{y}_2)s_{k_1}^3(\vec{y}_3) s_{k_4}^4(\vec{y}_4)+...\right].\nonumber
\eea
Here the dots stand for the other 22 terms. Now, one differentiates over the boundary values of the fields $s^1_{k_1}(\vec{x}_1),...,s^4_{k_4}(\vec{x}_4)$ and gets
\bea
\label{eq:exD}
	&&g_{1234}\,k_3k_4\left( D_{k_1 k_2 k_3 k_4}-2|\vec{x}_{34}|^2D_{k_1 k_2 k_3+1 k_4+1}\right) \nonumber\\
	&&~~~+ g_{3214}\,k_1k_4\left( D_{k_1 k_2 k_3 k_4}-2|\vec{x}_{14}|^2D_{k_1+1 k_2 k_3 k_4+1}\right)+...
\eea
Thus, in practice one can avoid writing down the relevant part of the contact Lagrangian altogether and write its contribution to the correlator directly, by simply finding the couplings corresponding to the 24 permutations of the weights and multiplying them by a correspondingly permuted set of $D$ functions as in \eqref{eq:exD}. This will give the contribution to the four-point function from the contact terms up to a normalization factor, which will be taken into account after computing the exchange part.
\subsection{Exchange term}
\label{sec:Exch}
The standard method of computing the exchange part, see e.g. \cite{Arutyunov:2000py}, is again to evaluate quadratic and cubic terms on-shell and differentiate over the boundary values of the fields. For this, one solves the equations of motion perturbatively and substitutes the solution into the full exchange Lagrangian. Writing down the latter already becomes unpleasant for higher-weight cases because of the growing number of descendants coupled to the scalars in the cubic terms. However, as for the contact part, the direct procedure can be streamlined and the contribution to the {\em non-normalized} four-point correlation function, coming from quadratic and cubic terms, can be written as the following sum of the $s$, $t$ and $u$ channels:
\bea
\label{eq:LExchange}
	\langle k_1 k_2 k_3 k_4 \rangle_{\text{Exchange}}&=&\sum \left[{36\ov \zeta(s_k)}S^{I_1I_2I_k}S^{I_3I_4I_k}\mathbf{S}^k_{k_1k_2k_3k_4} + {1\ov \zeta(A_{\m,k})}A^{I_1I_2I_k}A^{I_3I_4I_k}\mathbf{V}^k_{k_1k_2k_3k_4} \right.\nonumber\\
	&+& \left.{4\ov\zeta(\varphi_{\m\nu,k})}G^{I_1I_2I_k}G^{I_3I_4I_k}\mathbf{T}^k_{k_1k_2k_3k_4}+\ldots\right] \nonumber\\
	&+&\sum \left[{36\ov \zeta(s_k)}S^{I_1I_3I_k}S^{I_2I_4I_k}\mathbf{S}^k_{k_1k_3k_2k_4} + {1\ov \zeta(A_{\m,k})}A^{I_1I_3I_k}A^{I_2I_4I_k}\mathbf{V}^k_{k_1k_3k_2k_4} \right.\nonumber\\
	&+& \left.{4\ov\zeta(\varphi_{\m\nu,k})}G^{I_1I_3I_k}G^{I_2I_4I_k}\mathbf{T}^k_{k_1k_3k_2k_4}+\ldots\right] \nonumber\\
	&+&\sum \left[{36\ov \zeta(s_k)}S^{I_1I_4I_k}S^{I_2I_3I_k}\mathbf{S}^k_{k_1k_4k_2k_3} + {1\ov \zeta(A_{\m,k})}A^{I_1I_4I_k}A^{I_2I_3I_k}\mathbf{V}^k_{k_1k_4k_2k_3} \right.\nonumber\\
	&+& \left.{4\ov\zeta(\varphi_{\m\nu,k})}G^{I_1I_4I_k}G^{I_2I_3I_k}\mathbf{T}^k_{k_1k_4k_2k_3}+\ldots\right]. 
\eea
Here the exchange Witten diagrams $\mathbf{S}$, $\mathbf{V}$ and $\mathbf{T}$ are multiplied by the corresponding combination of quadratic and qubic couplings as follows from example \eqref{eq:eomL2}. 
\begin{table}[t]
\begin{center}
\begin{tabular}{ |c|c|c|c|c|c|c| } 
 \hline
 Field & $s_k$ & $A_{\m,k}$ & $C_{\m,k}$ & $\p_k$ & $t_k$ & $\varphi_{\m\nu,k}$\\ 
 \hline
 Irrep & $[0,k,0]$ & $[1,k-2,1]$ & $[1,k-4,1]$ & $[2,k-4,2]$ & $[0,k-4,0]$ & $[0,k-2,0]$\\ 
 \hline 
 $m^2$ & $k(k-4)$ & $k(k-2)$ & $k(k+2)$ & $k^2-4$ & $k(k+4)$ & $k^2-4$\\ 
 \hline
 $\De$ & $k$ & $k+1$ & $k+3$ & $k+2$ & $k+4$ & $k+2$\\ 
 \hline
\end{tabular}\caption{KK-modes contributing to the exchange Witten diagrams.}\label{tab:fields}
\end{center}
\end{table}
The dots indicate that we have only included the contribution of the fields $s_k$, $A_{\mu,k}$ and $\varphi_{\mu\nu,k}$ which are relevant for the computations central in this paper. The exchange contributions of the other three fields (see table \ref{tab:fields}) can easily be written down by simply changing the couplings and masses in the Witten diagrams. We sum over all possible Witten diagrams and the summation convention is the following: $$\sum\equiv \sum\limits_{k\in \{\text{Exch.\,fields}\}} \sum\limits_{I_k=1}^{\text{dim}([0,k,0])}.$$ Note also that the permutation of the weights $\{k_1,k_2,k_3,k_4\}$ takes place together with the permutation of the coordinates, e.g. $\mathbf{S}^k_{k_1k_3k_2k_4}\equiv\mathbf{S}^k_{k_1k_3k_2k_4}(\vec{x}_1,\vec{x}_3,\vec{x}_2,\vec{x}_4)$. The exchange fields which can show up in an exchange diagram are restricted by the SU$(4)$ selection rule, namely, these are the fields which appear in the intersection of non-vanishing cubic vertices, see table \ref{tab:fields}. This is determined by the following tensor product decomposition of representations:
\bea
	[0,k_1,0]\otimes[0,k_2,0]=\sum\limits_{r=0}^{\min(k_1,k_2)}\sum\limits_{s=0}^{\min(k_1,k_2)-r}[r,|k_1-k_2|+2s,r].
\eea
An exchange field must occur on the right-hand side of this decomposition for both the ingoing as well as the outgoing fields. 

The exchange Witten diagrams can be expressed as \emph{exchange integrals}, which are defined as follows:
{\small
\bea\label{eq:ExchInt}
	&&\mathbf{S}^k_{k_1k_2k_3k_4}(\vec{x}_1,\vec{x}_2,\vec{x}_3,\vec{x}_4)=\int[dz][dw]\,K_{k_1}(z,\vec{x}_1)K_{k_2}(z,\vec{x}_2)G^k(z,w)K_{k_3}(w,\vec{x}_3)K_{k_4}(w,\vec{x}_4),\nonumber\\
	&&\mathbf{V}^k_{k_1k_2k_3k_4}(\vec{x}_1,\vec{x}_2,\vec{x}_3,\vec{x}_4)=\nonumber\\
	&&~~~~~~~~~~=\int[dz][dw]\,K_{k_1}(z,\vec{x}_1)\overleftrightarrow{\na}^\m K_{k_2}(z,\vec{x}_2)G^k_{\m\nu}(z,w)K_{k_3}(w,\vec{x}_3)\overleftrightarrow{\na}^\nu K_{k_4}(w,\vec{x}_4),\nonumber\\
	&&\mathbf{T}^k_{k_1k_2k_3k_4}(\vec{x}_1,\vec{x}_2,\vec{x}_3,\vec{x}_4)=\int[dz][dw]\, T^{\m\nu}(z,\vec{x}_1,\vec{x}_2)G^k_{\m\nu\m'\nu'}(z,w)T^{\m'\nu'}(z,\vec{x}_3,\vec{x}_4),
\eea
}where $G^k$, $G^k_{\m\nu}$ and $G^k_{\m\nu\m'\nu'}$ are the scalar, vector and the tensor bulk-to-bulk propagators correspondingly. A simple method to compute the exchange integrals was introduced in \citep{DHoker:1999mqo} and further generalizations appeared in \citep{Arutyunov:2002fh} and \cite{Berdichevsky:2007xd}. As it turns out, they can always be represented as a finite sum of $D$ functions.

\begin{figure}[t]
\centering
\includegraphics[scale=1]{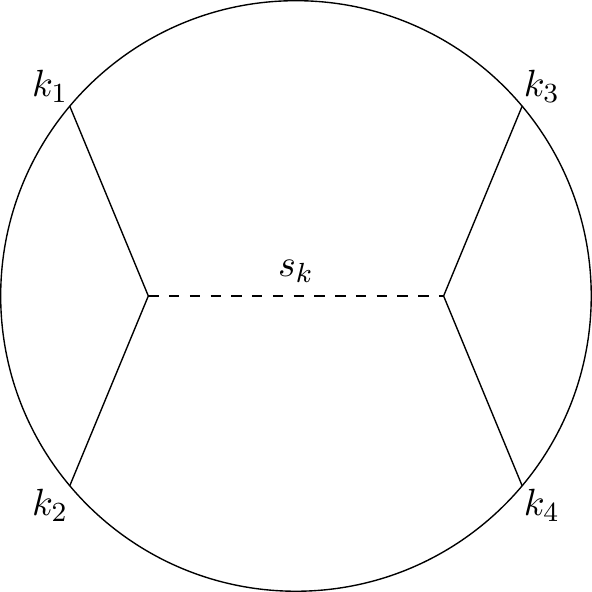}
\caption{The exchange Witten diagram where a scalar of weight $k$ is exchanged in the $s$ channel.}
\label{fig:Witten}
\end{figure}

The proof of \eqref{eq:LExchange} is very similar for all the six exchange fields in table \ref{tab:fields} for all the channels. Therefore we restrict ourselves and only show the derivation for a scalar field  of weight $k$ in the $s$ channel in detail here (see the Witten diagram in fig. \ref{fig:Witten}), following \cite{Arutyunov:2000py} closely.
The quadratic and cubic terms can be found in \cite{Arutyunov:1999fb} and are as follows\footnote{In order not to obstruct the reading, we omit the representation index $I_k$ and simply write $s_k \equiv s^{I_k}_k$. However, in the expressions similar to that one of $\mathcal{L}_3$, the usual summation convention as for \eqref{eq:summationconvention} is assumed.}: 
\bea
\label{eq:eomL2}
	\mathcal{L}_2^{(s)}&=&\sum_k\z(s_k)\left( -{1\ov 2}\na_\m s_k \na^\m s_k-{1\ov 2}m_k^2s_k^2\right),\nonumber\\
	\mathcal{L}_3^{(s)}&=&S_{I_1I_2I_3}s^{I_1}s^{I_2}s^{I_3}.
\eea
The equations of motion for scalars are
\bea
	\z(s_k)(\na_\m^2-m_k^2)\,s_k=-{\pa\mathcal{L}_3^{(s)}\ov \pa s_k}.
\eea
As was mentioned in section \ref{sec:Contact}, we are solving the equations of motion perturbatively: $s_k=\bar{s}_k+\tilde{s}_k$, where
\bea
	(\na_\m^2-m_k^2)\,\bar{s}_k&=&0,\nonumber\\
	(\na_\m^2-m_k^2)\,\tilde{s}_k&=&\left.-{1\ov \z(s_k)}{\pa\mathcal{L}^{(s)}_3\ov \pa s_k}\right|_{s\to\bar{s}}.
\eea    
Introducing the bulk-to-bulk propagator $G^k$ as the solution of: 
\bea
	(\na_\m^2-m_k^2)\,G^k(z,w)=-\de(z,w),
\eea
one gets
\bea
	\tilde{s}_k(z)={1\ov \z(s_k)}\int[dw]\left(\left.{\pa\mathcal{L}_3^{(s)}(w)\ov \pa s_k}\right|_{s\to\bar{s}}\right)G^k(z,w).
\eea
Using the equations of motion one finds the contribution to the four-point function coming from the quadratic term:
\bea
	\mathcal{L}_2=-{1\ov 2}\left(\left.{\pa\mathcal{L}^{(s)}_3\ov \pa s_k}\right|_{s\to\bar{s}}\right)\tilde{s}_k.
\eea
The cubic term, contributing to the four-point correlator, must be of the form $\bar{s}\bar{s}\tilde{s}$ and can be computed as follows:
\bea
	\mathcal{L}_3=\left(\left.{\pa\mathcal{L}^{(s)}_3\ov \pa s_k}\right|_{s\to\bar{s}}\right)\tilde{s}_k.
\eea
In this way one obtains the exchange part of the action:
\bea
	S_{\text{Exchange}}(s_k)= {1\ov 2\z(s_k)}\int[dz][dw]\left(\left.{\pa\mathcal{L}^{(s)}_3(z)\ov \pa s_k}\right|_{s\to\bar{s}}\right)G^k(z,w)\left(\left.{\pa\mathcal{L}^{(s)}_3(w)\ov \pa s_k}\right|_{s\to\bar{s}}\right).\nonumber\\
\eea
Writing down the cubic terms of interest, containing the exchange scalar $s_k$, explicitly
\bea
	\mathcal{L}^{(s)}_3=6S_{I_1I_2I_k}s_{k_1}^{I_1}s_{k_2}^{I_2}s_{k}^{I_k} + 6S_{I_3I_4I_k}s_{k_3}^{I_3}s_{k_4}^{I_4}s_{k}^{I_k},
\eea
we find
\bea
	S_{\text{Exchange}}(s_k)&=&\sum_k {1\ov 2\z(s_k)}\int[dz][dw]\left[6S_{I_1I_2I_k}\bar{s}_{k_1}^{I_1}\bar{s}_{k_2}^{I_2} + 6S_{I_3I_4I_k}\bar{s}_{k_3}^{I_3}\bar{s}_{k_4}^{I_4}\right](z)\cdot G^k(z,w)\nonumber\\
	&&~~~~~~~~~~~~~~~~~~~~~ \cdot\left[6S_{I_1I_2I_k}\bar{s}_{k_1}^{I_1}\bar{s}_{k_2}^{I_2} + 6S_{I_3I_4I_k}\bar{s}_{k_3}^{I_3}\bar{s}_{k_4}^{I_4}\right](w).\nonumber\\
\eea
Finally, using the solution \eqref{eq:boundsol} of the boundary problem and taking variational derivatives, we obtain the first term in \eqref{eq:LExchange}.

\subsection{Normalization}
The final step in computing the \emph{normalized} four-point function is to multiply the sum of the above-mentioned contact part \eqref{eq:fullcontact} and exchange part \eqref{eq:LExchange} by the following overall coefficient:
\bal
	\frac{N^2}{8\pi^5} &\prod\limits_{k\in\{ k_1,...,k_4\}} C_{O_k} C_k, \\
	C_{O_2}=\frac{4 \sqrt{2} \pi ^2}{N}, \quad  &C_{O_k}=\frac{2 \sqrt{2} \pi ^2}{N(k-2)\sqrt{ (k-1)}},~~k>2,
\eal
where $C_k$ are given in \eqref{eq:coeffCk} and the coefficients $C_{O_k}$ provide the canonical normalization for the two-point functions  \citep{Arutyunov:2000py}, \cite{Freedman:1998tz}. Note also that the minus sign coming from the  Euclidean version of the \AdS\, action compensates the minus sign in \eqref{eq:corrWitten}.
%
%
\section{Results}
\label{sec:results}
Using the simplified algorithm explained in the previous section we have computed the correlators $\langle 2345 \rangle$ and $\langle 3456 \rangle$, being the first-ever computed examples of all-different weights correlators and hence have no remaining symmetry in the outgoing legs. In addition, the $\langle 3456 \rangle$ is next-next-next-to-extremal, representing a really generic case. The only known example of these correlators thus far is the $\langle 5555 \rangle$ four-point function computed through the bootstrap approach in \cite{Rastelli:2017udc}, but this correlator is maximally symmetric in its external legs. 

Following section \ref{sec:generalities} we can write the correlators as a sum of a free and an interacting part, the latter of which can be expressed as in \eqref{eq:interacting}. We discuss the free parts further in section \ref{sec:extendedness} and list them in appendix \ref{app:intermediate results}. The conformally-invariant $\bar{D}$ functions appearing below follow from the $D$ functions defined in \eqref{eq:Dfunctions} by the formula
\begin{equation}
\bar{D}_{k_1,k_2,k_3,k_4}(u,v) = \frac{2}{\pi^2}\frac{\prod_{i=1}^4 \Gamma(k_i)}{ \Gamma\left( \Sigma-2\right)} \frac{x_{13}^{2\Sigma-2k_4} x_{24}^{2k_2}}{x_{14}^{2\Sigma-2k_1-2k_4}x_{34}^{2\Sigma-2k_3-2k_4}} D_{k_1,k_2,k_3,k_4}, \quad \text{with } \Sigma = \tfrac{1}{2} \sum_{i=1}^4 k_i.
\end{equation}
The tensors $T_l$ in \eqref{eq:Tl} depend implicitly on the chosen external legs and for the cases discussed here are explicitly given by 
\begin{equation}
\label{eq:2345T}
\begin{aligned}
T_1 &=t_{14}^2 t_{23}^2 t_{24} t_{34}^2, \quad &T_2 &= t_{13} t_{14} t_{23} t_{24}^2 t_{34}^2, \quad &T_3 &= t_{13}^2 t_{24}^3 t_{34}^2, \\
T_4 &=t_{12} t_{14} t_{23} t_{24} t_{34}^3, \quad &T_5 &=t_{12} t_{13} t_{24}^2 t_{34}^3 , \quad &T_6 &= t_{12}^2 t_{24} t_{34}^4,
\end{aligned}
\end{equation}
for $\langle 2345 \rangle$ and 
\begin{equation}
\label{eq:3456T}
\begin{aligned}
T_1 &=t_{14}^3 t_{23}^3 t_{24} t_{34}^2, \,\,\,\, &T_2 =& t_{13} t_{14}^2 t_{23}^2 t_{24}^2 t_{34}^2, \,\,\,\, &T_3 =&t_{13}^2 t_{14} t_{23} t_{24}^3 t_{34}^2 , \\
T_4 &=t_{13}^3 t_{24}^4 t_{34}^2, \,\,\,\, &T_5 =&t_{12} t_{14}^2 t_{23}^2 t_{24} t_{34}^3 , \,\,\,\, &T_6 =& t_{12} t_{13} t_{14} t_{23} t_{24}^2 t_{34}^3, \\
T_7 &= t_{12} t_{13}^2 t_{24}^3 t_{34}^3, \,\,\,\, &T_8 =& t_{12}^2 t_{14} t_{23} t_{24} t_{34}^4, \,\,\,\, &T_9 =& t_{12}^2 t_{13} t_{24}^2 t_{34}^4, \,\,\,\, \qquad T_{10} = t_{12}^3 t_{24} t_{34}^5,
\end{aligned}
\end{equation}
for $\langle 3456 \rangle$. In principle these are all the ingredients necessary to present the two correlators, however, their raw form as it is computed from the supergravity Lagrangian is quite complicated. 

\paragraph{$\langle 2345 \rangle$} ${}$ \\
Following eqn. \eqref{eq:interacting} the full \emph{interacting} part of the $\langle 2345 \rangle$ correlator depends on only a single $\mF$ function as
\bal
\label{eq:2345interacting}
	\langle 2345 \rangle_{\text{int}}  = {\sqrt{30}\ov N^2} \frac{x_{12}^4}{x_{13}^8 x_{24}^{10}}\frac{\mathcal{F}(u,v)}{u^3 v} \Big(&u\,T_1 + u (u-v-1)\,T_2 + u v\,T_3 + (-u-v+1)\,T_4 \\
	&+ v (-u+v-1)\,T_5 + v\,T_6  \Big)\,,
\eal
where $\mF$ is given by

\begin{equation}
\label{eq:2345result}
\mF = -\bar{D}_{2,3,4,7}. 
\end{equation}
This stunningly simple result follows after massaging the raw expression coming from the supergravity action considerably. In general one can simplify the raw result firstly by focussing on the large fractions appearing and showing that together they vanish due to identities for the $\bar{D}$ functions (see e.g. \cite{Uruchurtu:2008kp}). We were able to further simplify the resulting expression using knowledge of the simplified form of its Mellin transform: this transform suggests an ansatz for the correlation functions containing a small set of $\bar{D}$ functions, which ultimately leads to \eqref{eq:2345result}. We quote the raw and intermediate results in appendix \ref{app:intermediate results}. 

\paragraph{$\langle 3456 \rangle$} ${}$ \\
For $\langle 3456 \rangle$ a similar logic can be applied: the interacting correlator has the form
\begin{equation}
\label{eq:3456interacting}
\begin{aligned}
	\langle 3456 \rangle_{\text{int}} &= \frac{\sqrt{10}}{N^2} \frac{x_{12}^4}{x_{13}^{10} x_{24}^{12}} \frac{1}{u^4 v^2} \Big( u^2\mathcal{F}_1\, T_1 + u^2 \left( (u-v-1)\mathcal{F}_1+v\mathcal{F}_2 \right)\,T_2 \\
	&+ u^2 v \left(\mathcal{F}_1+ (u-v-1)\mathcal{F}_2\right)\, T_3 + u^2 v^2 \mathcal{F}_2\, T_4 + u \left(v\mathcal{F}_3 - (u+v-1)\mathcal{F}_1\right)\,T_5\\
	& -u v \left( (u-v+1)\mathcal{F}_1+ (u+v-1)\mathcal{F}_2+ (-u+v+1)\mathcal{F}_3\right)\,T_6 \\
	& + u v^2 \left( (-u+v-1)\mathcal{F}_2+\mathcal{F}_3\right)\,T_7 + v \left(u\mathcal{F}_1 - (u+v-1)\mathcal{F}_3\right)\,T_8\\
	& + v^2 \left((-u+v-1)\mathcal{F}_3 + u\mathcal{F}_2 \right)\,T_9 +  v^2 \mathcal{F}_3\,T_{10}\Big)\,,
\end{aligned}
\end{equation}
which can be written more compactly as 
\begin{equation}
	\langle 3456 \rangle_{\text{int}} = \frac{\sqrt{10}}{N^2}\frac{\mathcal{R}_{1,2,3,4}\, t_{24}t_{34}^2}{x_{12}^2x_{14}^2x_{23}^2x_{24}^2x_{34}^6} \left( u\, t_{14}t_{23} \mathcal{F}_1 + uv\, t_{13}t_{24} \mathcal{F}_2 + v\, t_{12}t_{34} \mathcal{F}_3\right). 	
\end{equation}
Thus, we only need to specify the three dynamical functions $\mF_i$. Their raw form as it follows from a direct computation is again complicated and we refer the interested reader to appendix \ref{app:intermediate results}. After following a similar approach as in the previous case the $\mF$ functions take the form
\begin{equation}
\label{eq:3456result}
\begin{aligned}
	\mathcal{F}_1 &= -\frac{1}{8} v (6 u+28 v+1) \bar{D}_{2,5,6,7}-\frac{5}{4} v \bar{D}_{2,4,6,8}-\frac{7}{8} v \bar{D}_{2,4,7,7}-\frac{5}{8} v \bar{D}_{2,5,5,8} \,, \\
	\mathcal{F}_2 &= -\frac{3}{2} u v \bar{D}_{2,6,5,7}-\frac{3}{2} v \bar{D}_{2,5,5,8}-3 v \bar{D}_{2,5,6,7}\,,\\
	\mathcal{F}_3 &= -\frac{7}{4} u v^2 \bar{D}_{2,6,6,6}-\frac{1}{8} v (10 u+12 v+3) \bar{D}_{2,5,6,7}-\frac{5}{8} u v \bar{D}_{2,6,5,7}-\frac{15}{8} v^2 \bar{D}_{2,5,7,6} \,. 
\end{aligned}
\end{equation} 
\section{Analyzing the results}
\label{sec:analyzing}
The direct results for the correlators $\langle 2345 \rangle$ and $\langle 3456 \rangle$ are quite complicated, warranting the need for good checks of the result. Apart from checking the expressions for the coupling building blocks known as $a$, $p$ and $t$ tensors using the identities in appendix \ref{app:C-algebra}, the most important check of the final result is verifying consistency with the structure explained in section \ref{sec:generalities}. 

In order to do this we need to separately compute the free part of the correlator $\langle k_1k_2k_3k_4\rangle_0$. In principle this computation is straightforward, but a subtlety concerning the identification of field theory operators with supergravity fields requires a discussion. We will address this subtlety -- due to the presence of \emph{extended operators} -- in section \ref{sec:extendedness}.

Assuming we have computed $\langle k_1k_2k_3k_4\rangle_0$ we can find the interacting part of the correlator. For the interacting part we can check whether it obeys the structure given in \eqref{eq:interacting}. This goes as follows: suppose the interacting part is provided as
\begin{equation}
\label{eq:aTstructure}
\sum_{l=1}^{|\{a\}|} \alpha_{l}\, T_l
\end{equation}
and we write the structure in \eqref{eq:interacting} as
\begin{equation}
\label{eq:bTstructure}
\sum_{m=1}^{|\{a\}|} \beta_{m}\, T_m,
\end{equation}
where the $\beta_m$ follow explicitly from \eqref{eq:interacting} and depend linearly on the $\mF_b$. Since we know that the number of independent functions describing the correlator is $|\{b\}|$, which is always strictly smaller than $|\{a\}|$, equating the expressions \eqref{eq:aTstructure} and \eqref{eq:bTstructure} yields an overdetermined system of equations for the unknowns $\mF_b$. The fact that our results satisfy this system provides a highly non-trivial check of the correlators. 

In fact, quite similarly to the coordinate space method from \cite{Rastelli:2017udc}, this system allows one to determine many of the numbers one has to compute during the computation. For example, in all our examples this provided an independent check of the symmetry factors in \eqref{eq:LExchange} and with a small modification we used it to verify the free part of our correlators. 

We can further analyze our results by comparing it to the closed Mellin-space formula that was conjectured for any four-point function of $1/2$ BPS operators \cite{Rastelli:2016nze} (see eqn. 25 in \cite{Rastelli:2016nze}). This conjecture is based on physical arguments and on consistency with all the known results and subsequently tested on one new result (the $\langle 5555 \rangle$ correlator) \cite{Rastelli:2017udc}. The correlators presented in section \ref{sec:results} therefore provide a new test of this conjecture, in particular because of their genericness. 

We have performed this test as follows: the conjecture for the correlator is given explicitly for the Mellin transform of the function $\mathcal{H}$ in the decomposition 
\begin{equation}
\label{eq:RHdecomp}
\mathcal{G}_{\text{conn}} = \mathcal{G}_{0, \text{conn}} + R \,  \mathcal{H},
\end{equation}
where $\mathcal{G}_{\text{conn}}$ is the conformal connected correlator obtained from the full correlator by subtracting the disconnected graphs and rescaling the result by a rational function of the coordinates\footnote{Note in addition that in \cite{Rastelli:2016nze} the weights are ordered $p_1 \geq p_2 \geq p_3 \geq p_4$, exactly opposite to our conventions. So matching the conjecture includes reshuffling the weights as well.}. Furthermore, $\mathcal{G}_{0, \text{conn}}$ is the rescaled connected free part and $R$ is a fixed function of conformal cross-ratios and inner products of SO$(6)$ vectors similar to \eqref{eq:Rpolynomial}. To check the conjecture we therefore first need to decompose our results in the form \eqref{eq:RHdecomp}: this can be done by solving a set of linear equations obtained from the decomposition into different tensor components. The resulting expressions for $\mathcal{H}$ are expressed as a linear combination of $\bar{D}$ functions and can then be straightforwardly Mellin-transformed using the Mellin-Barnes representation for $\bar{D}$ functions:
\begin{equation}
\label{eq:MellinDb}
\begin{aligned}
\bar{D}_{\Delta_1, \ldots, \Delta_4} (u,v) &= 2 \int \frac{ds}{2}\frac{dt}{2} u^{\tfrac{s}{2} - \frac{\Delta_1 +\Delta_2}{2}} v^{\tfrac{t}{2} - \frac{\Delta_2 +\Delta_3}{2}}
\Gamma\left(\tfrac{-s+\Delta_1+\Delta_2}{2} \right)
\Gamma\left(\tfrac{-s+\Delta_3+\Delta_4}{2} \right)\times \\
&\Gamma\left(\tfrac{-t+\Delta_1+\Delta_4}{2} \right)
\Gamma\left(\tfrac{-t+\Delta_2+\Delta_3}{2} \right)
\Gamma\left(\tfrac{s+t-\Delta_2-\Delta_4}{2} \right)
\Gamma\left(\tfrac{s+t-\Delta_1-\Delta_3}{2} \right),
\end{aligned}
\end{equation}
where $s$ and $t$ are the Mandelstam variables that are the coordinates in Mellin space. The resulting Mellin-space expressions are rational functions in these coordinates and further simplification of them yields an exact match with the conjecture up to an expected normalization constant $f$ that was found in \cite{Aprile:2018efk}. This therefore corroborates the conjecture. 

Alternatively, one can also check the Mellin-space conjecture by directly Mellin transforming the connected correlator $\mathcal{G}_{\text{conn}}$. This can then subsequently be matched to a different Mellin-space expression that one obtains from the conjecture by acting with a difference operator that one can build from $R$ (see eqn. 4.27 in \cite{Rastelli:2017udc}). We also performed this check for both our cases and obtained a match between the expressions coming from our results and those from the conjecture. 
\section{Computing the free part: extended operators}
\label{sec:extendedness}
In principle the computation of the free part is a straightforward procedure. We can compute it in the field theory picture using the (free) operators dual to the scalar fields $s_k^{I_k}$, simply performing the relevant Wick contractions to obtain the free correlator. For small $k<4$ the correspondence between fields and operators is simple, namely $s_k^{I_k} \sim \mathcal{O}^{I_k}_k$ where, remembering the decomposition \eqref{eq:BPSdecomp}, the field dependence is given by
\be
\mathcal{O}^{i_1\ldots i_k} = \kappa_k \text{Tr}\left(\phi^{i_1}\ldots \phi^{i_k}\right),
\ee
where $\kappa_k$ is the $k$-dependent normalization determined by demanding canonical two-point functions which in the planar limit can be taken to be $\kappa_k = \sqrt{2^k/(k N^k)}$. The operators with this exact field-dependence are known as single-trace CPO's and we distinguish them by omitting the tilde. However, as first noticed in \cite{DHoker:1998ecp} and later further analyzed in \cite{Arutyunov:2000ima} this correspondence cannot hold when $k\geq 4$: the fact that extremal three-point functions $\langle s_{k_1}^{I_1} s_{k_2}^{I_2} s_{k_3}^{I_3} \rangle$ vanish when computed from the supergravity Lagrangian, whereas the quantity
\be
\langle \mO^{I_1}_{k_1} \mO^{I_2}_{k_2} \mO^{I_3}_{k_3} \rangle
\ee
in the field theory in general is non-vanishing shows that this correspondence cannot continue to hold. The resolution presented in \cite{Arutyunov:2000ima} is that the scalar fields are not dual to single-trace CPO's, but to so-called \emph{extended} CPO's\footnote{One could alternatively leave the field operators unaltered and modify the supergravity Lagrangian instead, by a field redefinition that adds boundary terms such that the three-point correlation functions reflect the field theory result \cite{Arutyunov:2000ima}.}: 
\be
s_{k_1}^{I_1} \sim \tilde{\mO}^{I_1}_{k_1} = \mO^{I_1}_{k_1} - \frac{1}{2 N} \sum _{\substack{k_2,k_3\geq 2 \\
k_2 + k_3 = k_1}} C^{I_1 I_2 I_3} \mO^{I_2}_{k_2} \mO^{I_3}_{k_3},
\ee
where $C^{I_1 I_2 I_3} = \sqrt{k_1 k_2 k_3} \langle C^{I_1} C^{I_2} C^{I_3} \rangle$, defined in appendix \ref{app:C-algebra}. For convenience we also give the decomposition \eqref{eq:BPSdecomp} for the extended case: decomposing as in \eqref{eq:BPSdecomp} 
we can write the field-dependence as
\be
\tilde{\mO}^{i_1,\ldots, i_{k_1}} = \kappa_{k_1} \text{Tr}\left( \phi^{i_1}\ldots \phi^{i_{k_1}}\right) -
\sum_{\substack{k_2,k_3\geq 2 \\ k_2 + k_3 =k_1}} \frac{\lambda_{k_1,k_2,k_3}}{2N}\kappa_{k_2}\kappa_{k_3} \text{Tr}\left( \phi^{i_1}\ldots \phi^{i_{k_2}}\right) \text{Tr}\left( \phi^{i_{k_2+1}}\ldots \phi^{i_{k_1}}\right),
\ee 
where $\lambda_{k_1,k_2,k_3} = C^{I_1 I_2 I_3}$ and where we omitted symmetrization over all the indices since this is enforced by contraction with the $t$ vectors. Even though the prescription for $\tilde{\mO}^{I_1}_{k_1}$ is completely explicit it can be quite non-trivial to compute the coefficient directly, due to the complexity of the required tensor contractions. Luckily we can circumvent this by noting that the extended CPO's are required to have vanishing extremal three point functions. Since the number of terms in the summation is exactly equal to the number of extremal three-point functions containing $\tilde{\mO}^{I_k}$ one can find the coefficients by demanding vanishing of these three-point functions. For example, we can list the values of the $\lambda$ for the first few cases:
\be
\lambda_{422} = 4, \quad \lambda_{532} = 2 \sqrt{30}, \quad \lambda_{642} = 8 \sqrt{3}, \quad \lambda_{633} = 3 \sqrt{6},
\ee
thereby completely defining the extended operators dual to $s_4, s_5$ and $s_6$. 
\\[5mm]
The free part of the supergravity correlators should generically be computed using Wick contractions of the extended operators and then taking the large $N$ limit. However, the leading (planar) order in this computation follows from general considerations of the topology of the diagram combined with some combinatorics. In particular, from these considerations it follows that for connected diagrams the effect of the presence of extended operators was undetectable except for the extremal cases, such that in practice one did not have to consider this complication. Indeed, for the $\langle 4444 \rangle$ correlator, one of the few known correlators for which the weights are high enough to potentially feel this effect, the free part was computed in \cite{Arutyunov:2003ae} without explicitly tracking it. Consistency with superconformal symmetry was shown, thereby indicating that the fact that the operators are extended should not play a role. However, the first signs that one should take this effect seriously were presented in the paper \cite{Uruchurtu:2008kp}, that discusses the family of correlators of the form $\langle 22nn \rangle$ for $n\geq 2$: already there it was noted that there exists a discrepancy between the free part of the correlator as computed from supergravity as opposed to the result from field theory using non-extended operators when $n>3$, but its origin remained unexplained. It was argued in \cite{Rastelli:2017udc} that this discrepancy is resolved by computing the free part using extended operators. We confirm this with the explicit computation of the $\langle 22nn \rangle$ correlator for $n=4,5,6$, for which we conclude that the presence of extended operators does play a role. 

\begin{figure}
\includegraphics[width=\textwidth]{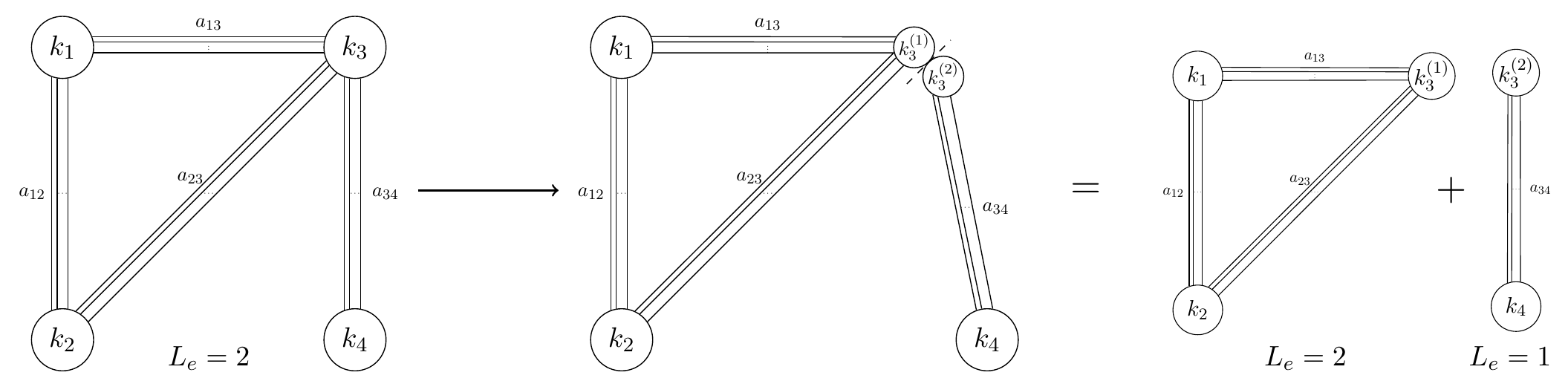}
\caption{An example of a singly-connected graph which has more index loops after splitting: the number of external index loops $L_e$ (loops outside of the lines that connect the nodes)  before splitting is $2$, whereas after splitting the total number is $2+1 = 3$.}
\label{fig:freepart}
\end{figure}

Moreover, since the computation only concerns the planar diagrams it is possible to prove when it is necessary to take into account that the operators are extended: in the planar limit only the leading order in $N$ of the free correlator described by a diagram is relevant. From basic observations it is known that, when depicted using the double-line notation, this leading order goes as $N^I$ with $I$ the number of index loops \cite{tHooft:1973alw}. Extendedness of an operator adds to a diagram a contribution of a second diagram in which one of the vertices has been split into two parts. As an example, splitting the third vertex looks like
\begin{equation}
A_{k_1, k_2, k_3, k_4} \rightarrow A_{k_1, k_2, k_3, k_4} - \frac{\#}{N}  A_{k_1, k_2, k_3^{(1)}, k_3^{(2)}, k_4},
\end{equation}
where $k_3^{(1)} + k_3^{(2)} = k_3$. In terms of the index loops, splitting the vertex will generically reduce the number of index loops by $1$, which when combined with the extra $1/N$ in front of the second diagram implies that its contribution is subleading and can be discarded. Only in the special case in which one of the vertices is singly-connected to the rest of the diagram and can be split off completely by one of the extensions -- hence yielding a disconnected diagram -- can the effect be leading: in that case the number of index loops increases by $1$ due to the splitting, which when compensated by the $1/N$ prefactor yields a contribution to the leading term and hence to the planar free correlator. We have illustrated these ideas in fig. \ref{fig:freepart}. 
\\[5mm]
In particular, it follows that the presence of extended operators plays a role for all the correlators presented in the previous section. For the more empirically inclined reader we provide an overview of the (extended and non-extended) free parts of all the non-trivial four-point functions with weights up to and including $5$ in appendix \ref{app:freeparts}. 

\section{Discussion}
\label{sec:discussion}
In this paper we have discussed simplifications to the algorithm that computes four-point functions of $1/2$ BPS operators from the effective supergravity Lagrangian, in particular presenting a closed formula for the exchange part \eqref{eq:LExchange} and further simplifying the contributions of the quartic Lagrangian (see eqn. \eqref{eq:fullcontact}). Combined with simplifications to the computation of the couplings in the contact Lagrangian (see appendix \ref{app:C-algebra}) these facilitate the direct computation of many new correlators. We have presented the computation of the all-different-weight four-point functions $\langle 2345 \rangle$ and $\langle 3456 \rangle$, whose simplest form can be found in \eqref{eq:2345result} and \eqref{eq:3456result} respectively. By Mellin-transforming these results we have shown that they match the recent conjecture \cite{Rastelli:2016nze,Rastelli:2017udc} for four-point functions. 

The simplifications of the algorithm presented here could be used to efficiently compute exactly many other correlators as well, as long as the weights are not too high: the only unknowns one needs to compute are the $a$, $p$ and $t$ tensors corresponding to a given set of weights. For example, the correlators $\langle 5667 \rangle$ and $\langle 4578 \rangle$ should be computable.  Another direction would be to further develop the coordinate space bootstrap program initiated in \cite{Rastelli:2017udc} using these simplifications. Also, it is worth investigating how the harmonic-polynomial formalism developed in \cite{Dolan:2003hv,Nirschl:2004pa} can be used for further simplifications. 

Alternatively, new correlators can be used to further develop the program that studies the strong-coupling OPE of the $\mN=4$ primary operators and help unravelling the mixing problem \cite{Alday:2017xua,Aprile:2017qoy,Aprile:2017xsp,Aprile:2018efk}. 

Finally, a more speculative direction is that of proving the Mellin-space conjecture directly from the supergravity Lagrangian by computing the four-point function for arbitrary weights. It is unclear how to do this, but the closed formula for the exchange part of the correlator is a step forwards in this direction. A further refinement of the algorithm to this aim, for example by finding and implementing additional reduction formulae (see appendix \ref{app:C-algebra}), is certainly possible. 

\section*{Acknowledgments} We would like to thank Till Bargheer, Vsevolod Chestnov and Enrico Olivucci for useful discussions. In particular we thank Sergey Frolov for collaborations during the initial stages of this work and feedback on the manuscript. This work is supported by the German Science Foundation (DFG) under the Collaborative Research Center (SFB) 676 ``Particles, Strings and the Early Universe" and the Research Training Group (RTG) 1670 ``Mathematics inspired by String Theory and Quantum Field Theory". 
\appendix
\section{C-algebra}\label{app:C-algebra}
\subsection{Definitions}
The $A$ and $S$ couplings in the Lagrangian from \eqref{eq:adsaction} are represented as sums of Clebsch-Gordon coefficients for SO$(6)$ irreps that come in three types: $a$, $t$ and $p$ tensors. These expressions where computed in \cite{Arutyunov:1999fb} and in principle describe all the couplings, but in practice their computation can be cumbersome. This is due to the fact that they contain contractions of $C$ tensors, which carry the tensor structure of the correlator as described in section \ref{sec:generalities}. Since each $C$ tensor carries roughly as many indices as the weight of the representation it belongs to this can get unwieldy quickly, explaining why until now almost no correlators are known with weights larger than four.\footnote{The only exceptions are the family $\langle 22 nn \rangle $ for $n>1$  in \cite{Uruchurtu:2008kp} and $\langle 5555 \rangle $ in \cite{Rastelli:2017udc}, which both feature a large degeneracy allowing for simplifications. } 
\\[5mm]
To be more precise, following the notation from \cite{Arutyunov:1999fb}, we want to compute products of the objects
\begin{equation}
\label{eq:aptintegrals}
a_{123}  = \int Y^{I_1}Y^{I_2}Y^{I_3}, \quad
t_{123}  = \int \nabla^{\a}Y^{I_1}Y^{I_2}Y_{\a}^{I_3}, \quad  
p_{123}  = \int \nabla^{\a}Y^{I_1}\nabla^{\b}Y^{I_2}Y_{\left(\a \b \right)}^{I_3},  
\end{equation}
where the $Y$ functions are spherical harmonics on the five-sphere and the round brackets indicate traceless symmetrization. More precisely, we need to find expressions for $a_{125}a_{345}$, $t_{125}t_{345}$ and $p_{125}p_{345}$, where $1,2,3,4$ refer to the weights $k_1,k_2,k_3,k_4$ of the correlator and $5$ is an intermediate leg with a weight $k_5$. The integrals \eqref{eq:aptintegrals} can be expressed using so-called $C$-tensors (see for more details \cite{Arutyunov:1999fb}). For example, $a_{123}$ is defined as
\begin{equation}
\label{eq:atensorexpression}
a_{123} = \frac{\prod_{i=1}^{3} \frac{k_i z(k_i)}{\alpha_i !}}{\pi^{\tfrac{3}{2}} (\sigma +2)! \, 2^{\sigma-1}} \left\langle C^{I_1}_{[0,k_1,0]}C^{I_2}_{[0,k_2,0]}C^{I_3}_{[0,k_3,0]} \right\rangle,
\end{equation}
where $\alpha_i = \alpha_{i}(k_1,k_2,k_3) = \sigma -k_i $ for $i=1,2,3$ and $\sigma = \sigma(k_1,k_2,k_3) = (k_1+k_2 +k_3)/2$ and where
\begin{equation}
\left\langle C^{I_1}_{[0,k_1,0]}C^{I_2}_{[0,k_2,0]}C^{I_3}_{[0,k_3,0]} \right\rangle = 
C^{I_1}_{i_1\ldots i_{\alpha_2} j_{1} \ldots j_{\alpha_3}} C^{I_2}_{j_1\ldots j_{\alpha_3} l_{1} \ldots l_{\alpha_1}}C^{I_3}_{l_1\ldots l_{\alpha_1} i_{1} \ldots i_{\alpha_2}}
\end{equation}
encodes the tensor structure of the correlator. The number of non-zero tensors is restricted by representation theory and therefore finite. 
\\[5mm]
Although the formulae for $a_{125}a_{345}$, $t_{125}t_{345}$ and $p_{125}p_{345}$ are fully explicit it is not straightforward to compute them. The main obstruction in performing this computation therefore in computing a correlation function from the Lagrangian sits in having to use completeness relations for the $C$ tensors: for all of the $a$, $t$ and $p$ tensors one has to evaluate objects of the form
\begin{equation}
\label{eq:CCproduct}
\sum_{I_5} \langle C^{I_1}_{k_1} C^{I_2}_{k_2}C^{I_5}_{k_5} \rangle \langle C^{I_3}_{k_3} C^{I_4}_{k_4}C^{I_5}_{k_5} \rangle,
\end{equation}
where the sum is over the representation index of the $k_5$ field. The completeness relations allow us to evaluate this sum, such that only four $C$ tensors remain that encode the tensor structure of the correlator. For $a$ tensors a closed formula for the completeness relation exists (given below in \eqref{eq:B5Arutyunov}), but its evaluation can become impossible for expressions for higher-weight operators due to computational limits. For $t$ and $p$ tensors no such formula exists at present and the best we can do is determine the completeness relations for a fixed set of weights from the properties of $C$ tensors. Additionally, we can use so-called reduction relations to reduce the total number of unknown $a$, $t$ and $p$ tensors. They are discussed in the next section \ref{sec:reduction}. These relations allow us to express the $t$ and $p$ tensors with the heighest intermediate $k_5$ weights in terms of $a$, $t$ and $p$ tensors we already know.
\\[5mm]
To summarize, the procedure to obtain the $a$, $t$ and $p$ tensors for a correlator with given weights $(k_1,k_2,k_3,k_4)$ is the following: we compute the necessary $a$ tensors using the formula \eqref{eq:atensorexpression} involving a completeness relation that we will get back to in more detail in the next section. Then we use the reduction relations discussed in section \ref{sec:reduction} to express the highest weight $t$ and $p$ tensors using lower ones and some of the $a$ tensors we computed. We compute the remaining necessary $t$ and $p$ tensors by explicitly finding the completeness relations for the given weights by imposing all the properties of $C$ tensors on an ansatz. This yields the complete set of needed $a$, $t$ and $p$ tensors. 

\subsection{Simplifying completeness relations for $a$ tensors}
To compute the $a$ tensors we need expressions for the completeness relation of $C$ tensors that allow us to resolve the sum in \eqref{eq:CCproduct}. Here we will focus only on simplification of the case for $a$ tensors, since there is an explicit formula available. Ultimately we need to reduce \eqref{eq:CCproduct} to a sum of the independent tensor structures of the form $C^{I_1}C^{I_2}C^{I_3}C^{I_4}$ that carry the tensor structure of the correlator only, meaning we should get rid of the sum over the representation index $I_5$. This can be done in principle as this sum constitutes a completeness condition for the $C$ tensors, meaning it is expressable as a linear combination of products of Kronecker delta functions carrying the indices, i.e.
\begin{equation}
\label{eq:completeness}
\sum_{I}C^I_{i_1 \ldots i_n}C^I_{i_{n+1} \ldots i_{2n}} = \sum_{\sigma \in S_{2n}}A_{\sigma} \delta_{i_{\sigma(1)}i_{\sigma(2)}}\delta_{i_{\sigma(3)}i_{\sigma(4)}}\ldots \delta_{i_{\sigma(2n-1)}i_{\sigma(2n)}},
\end{equation}
where $S_n$ is the symmetric group of $n$ objects and $A_{\sigma}$ are coefficients that are to be determined from the properties of the $C$ tensors: after taking into account the internal symmetries of the product of delta functions, being symmetric under exchange of indices belonging to the same Kronecker delta and permutation of these delta functions, the number of coefficients is $(2n)!/(2^n n!)$. For the main case $\langle 3456 \rangle$ under consideration in this paper the largest $n$ one has to consider is $n=9$, yielding over $34$ million terms. 
\\[5mm]
Although this is indeed the whopping number of coefficients that needs to be computed and stored to express the completeness condition \eqref{eq:completeness} we can consider a simplified version of it to compute the sum in formula \eqref{eq:CCproduct}, since the other $C$ tensors have additional sym metries. Our starting point to derive this simplification is the fully explicit formula given in B.5 of \cite{Arutyunov:2002fh}:
\begin{equation}
\label{eq:B5Arutyunov}
C_{i_1,\ldots, i_n}^IC_{j_1,\ldots, j_n}^I = \sum_{k=0}^{\left \lfloor{\tfrac{n}{2}}\right \rfloor } \theta_{k} \sum_{\{l_1\ldots l_{2k}\}} \delta_{i_{l_1}i_{l_2}}\ldots \delta_{i_{l_{2k-1}}i_{l_{2k}}}\tilde{\delta}^{(n-2k)}_{i_1\ldots \hat{i}_{l_1}\ldots \hat{i}_{l_{2k}}\ldots i_n,(j_{2k+1}\ldots j_n}
\delta_{j_1 j_2}\ldots \delta_{j_{2k-1}j_{2k})},
\end{equation}
where the sum over $\{l_1\ldots l_{2k}\}$ runs over all subsets of $\{1,\ldots, n\}$ containing $2k$ elements that yield inequivalent products of delta functions, $(\ldots)$ stands for symmetrization of the indices and we use the definition $\tilde{\delta}^{(p)}_{i_1,\ldots i_p, j_1,\ldots j_p} = \delta^{(p)}_{(i_1,\ldots i_p), (j_1,\ldots j_p)}$ and $\delta^{(p)}_{i_1,\ldots i_p, j_1,\ldots j_p} = \prod_{k=1}^p \delta_{i_k,j_k}$. The coefficients $\theta_k$ are known explicitly as 
\begin{equation}
\theta_0=1,\quad \text{if } k>0: \, \theta_k = \frac{(-1)^k}{2^k(n+1)\ldots(n+2-k)}
\end{equation}
However, the rapidly growing number of summands makes this expression difficult to work with for higher weights, also because it contains symmetrizations that lead to overcounting in the intermediate stages. Since all the $C$ tensors are traceless and completely symmetric in their indices we can restrict the formula further. In order to provide a succinct derivation we first discuss some theory related to a splitting of the symmetric group.
\subsubsection{Decomposing the symmetric group}
To simplify the completeness condition it will prove convenient to be able to decompose elements of the symmetric group $S_n$: consider two natural numbers $p,q$ such that $p+q=n$, then $S_p$ and $S_q$ are subgroups of $S_n$ but generically they will not be normal such that the quotient $S_n/S_p$ is not well-defined as a group. Nevertheless, it is possible to decompose any element $\sigma \in S_n$ as a product $\sigma = \rho \sigma_q \sigma_p$, where $\sigma_p$ only permutes the first $p$ objects, $\sigma_q$ only permutes the last $q$ objects and $\rho$ is a product of transpositions that each swap one object from the first $p$ with one of the last $q$. The only complication is that, in the cases in which $S_p$ and/or $S_q$ are not normal subgroups, the set of elements $\rho$, which we will denote by
\begin{equation}
 S_n \sslash \left(S_p\times S_q\right) \text{ or simply }  S_n^{(p,q)} 
\end{equation}
is not a subgroup as it is not even closed under multiplication. It is also not uniquely defined, as multiplication by an element from $S_p$ or $S_q$ gives a new set with the same properties. This is ultimately not problematic as any set of representatives will do to perform the sums we are interested in. We will present here one choice for this set.\footnote{It seems unlikely that the group theory, decomposition and choice of representatives we present here are all novel, but despite a considerable effort we have not managed to find other accounts. Since it helps to simplify the contraction of symmetric tensors one could nevertheless fathom a number of applications where it could help speed up computations. The authors would welcome seeing any application of these principles, for example an implementation into Form.}
\subsubsection{A characterization of  $S_n\sslash \left(S_p\times S_q\right)$}
First of all, the counting tells us that $S_n^{(p,q)} $ should have $\tfrac{n!}{p! q!}$ elements which should all be independent, i.e. should not be expressable as a product of other elements in the set. Secondly, since $S_p$ and $S_q$ already take care of all the rotations in the first $p$ and last $q$ objects we can consider only those permutations which are built up from transpositions that lie outside both $S_p$ and $S_q$. The remaining transpositions are necessarily of the form $(i j)$ with $1\leq i\leq p$ and $n\geq j>p$. Thirdly, if in a product of transpositions two of them contain the same number $i$ this can be taken away by a suitable transposition from either $S_p$ or $S_q$. From this we conclude that $S_n^{(p,q)} $ contains all the permutations that can be built from the transpositions $(i j)$ with $i\leq p$ and $j>p$ such that all of them are disjoint. Their number is easily computed:
\begin{itemize}
\item The identity transposition: $1$ element
\item All single transpositions: there are $pq$ of them
\item All products of two disjoint transpositions: there are $pq(p-1)(q-1)/(2!\cdot2!)$ of those
\item In general the products of $r$ disjoint transpositions are
\begin{equation}
\frac{\prod_{k=0}^{r-1}(p-k)(q-k)}{r! \cdot r!}
\end{equation}
in number.
\end{itemize}
This means that we have formed a set containing
\begin{equation}
\sum_{l=0}^{\text{min}(p,q)} \frac{1}{(l!)^2} \prod_{k=0}^{l-1}(p-k)(q-k) = \frac{(p+q)!}{p! q!}
\end{equation}
elements. Since the three sets $S_p$, $S_q$ and $S_n^{(p,q)} $ are all disjoint and independent we see that our characterization results in the correct number of elements $\sigma= \rho \sigma_q \sigma_p$, namely $n!$, implying that this description will work to decompose $\sigma$. Finally, we will use the notation $S_n^{(p,q)}(k)$ to denote the part of $S_n^{(p,q)}$ containing those elements built up from exactly $k$ transpositions. 
\subsubsection{Example}
As an example we present the set $S_7^{(4,3)} $, which contains the $35$ elements listed in the following table:
\\[5mm]
\begin{tabular}{ l | c | p{12cm} }
  part of $S_7^{(4,3)}$ & \# & elements  \\
  $S_7^{(4,3)}(0)$ & $1$ &\text{id} \\ \hline
  $S_7^{(4,3)}(1)$ & $12$  & $(1,5)$, $(1,6)$, $(1,7)$, $(2,5)$, $(2,6)$, $(2,7)$, $(3,5)$, $(3,6)$, $(3,7)$, $(4,5)$, $(4,6)$, $(4,7)$ \\  \hline
  $S_7^{(4,3)}(2)$& $18$ & $(1,5)(2,6)$, $(1,5)(2,7)$, $(1,5)(3,6)$, $(1,5)(3,7)$, $(1,5)(4,6)$, $(1,5)(4,7)$, $(1,6)(2,7)$, $(1,6)(3,7)$, $(1,6)(4,7)$, $(2,5)(3,6)$, $(2,5)(3,7)$, $(2,5)(4,6)$, $(2,5)(4,7)$, $(2,6)(3,7)$, $(2,6)(4,7)$, $(3,5)(4,6)$, $(3,5)(4,7)$, $(3,6)(4,7)$ \qquad\qquad \,  \\[-5mm]  \hline
  $S_7^{(4,3)}(3)$ & $4$ & $(1,5)(2,6)(3,7)$, $(1,5)(2,6)(4,7)$, $(1,5)(3,6)(4,7)$, $(2,5)(3,6)(4,7)$ \qquad \qquad\qquad
\end{tabular}
\subsubsection{Simplifying the completeness condition}
We can now simplify the completeness condition using the fact that the indices $\mathcal{I} = (i_1,\ldots i_n)$ and $\mathcal{J}=(j_1,\ldots j_n)$ are contracted with the indices of traceless symmetric $C$ tensors in the expression \eqref{eq:CCproduct}: explicitly we can write \eqref{eq:CCproduct} as
\begin{equation}
\sum_{I_5} \sum_{\mK_1, \mK_2} \sum_{\mI,\mJ} C_{\mI_2 \mK_1}^{I_1}C_{\mK_1 \mI_1}^{I_2} C_{\mI_1 \mI_2}^{I_5} C_{\mJ_2 \mK_2}^{I_3}C_{\mK_2 \mJ_1}^{I_4} C_{\mJ_1 \mJ_2}^{I_5},
\end{equation}
where $\mI = \mI_1 \cup \mI_2$ and $\mJ = \mJ_1 \cup \mJ_2$ are multi-indices that are summed over with constituents 
\begin{equation}
\begin{aligned}
\mI_1 &=\{i_1,\ldots, i_{\alpha_1(k_1,k_2,k_5)}\},\qquad &\mJ_1 &= \{j_1,\ldots, j_{\alpha_1(k_3,k_4,k_5)}\}, \\
\mI_2 &= \{i_{\alpha_2(k_1,k_2,k_5)+1},\ldots, i_{n}\}, \qquad &\mJ_2 &= \{j_{\alpha_2(k_3,k_4,k_5)+1},\ldots, j_{n}\},
\end{aligned}
\end{equation}
and $\mK_1, \mK_2$ are other multi-indices containing $\alpha_3(k_1,k_2,k_5)$ and $\alpha_3(k_3,k_4,k_5)$ indices respectively. In the following we will leave the sums implicit. We will now insert \eqref{eq:B5Arutyunov}:
\begin{equation}
\begin{aligned}
C_{\mI_2 \mK_1}^{I_1}C_{\mK_1 \mI_1}^{I_2}  C_{\mJ_2 \mK_2}^{I_3}C_{\mK_2 \mJ_1}^{I_4} C_{\mI_1 \mI_2}^{I_5}C_{\mJ_1 \mJ_2}^{I_5}= C_{\mI_2 \mK_1}^{I_1}C_{\mK_1 \mI_1}^{I_2}  C_{\mJ_2 \mK_2}^{I_3}C_{\mK_2 \mJ_1}^{I_4} \times  \\
\sum_{k=0}^{\left \lfloor{\tfrac{n}{2}}\right \rfloor } \theta_{k} \sum_{\{l_1\ldots l_{2k}\}} \delta_{i_{l_1}i_{l_2}}\ldots \delta_{i_{l_{2k-1}}i_{2k}}\tilde{\delta}^{(n-2k)}_{i_{l_1}\ldots \hat{i}_{l_1}\ldots \hat{i}_{l_{2k}}\ldots i_n,(j_{2k+1}\ldots j_n}
\delta_{j_1 j_2}\ldots \delta_{j_{2k-1}j_{2k})}.
\end{aligned}
\end{equation}
Let us first note that the first product of delta functions $\delta_{i_{l_1}i_{l_2}}\ldots \delta_{i_{l_{2k-1}}i_{2k}}$ is being contracted with two traceless tensors. This implies that whenever $l_r,l_s$ belong to the same index set $\mI_1$ or $\mI_2$ the contraction will make its contribution vanish. Interestingly, this implies that we can restrict the sum over subsets $\{l_1,\ldots, \l_{2k}\}$ further using our newly defined subsets $S_n^{(p,q)}(k)$. Let $\tau = (n_1 n_2)\ldots (n_{2k-1} n_{2k})$ be a product of $k$ transpositions, then we define 
\begin{equation}
\delta_{\tau} \coloneqq \delta_{n_1,n_2}\ldots \delta_{n_{2k-1} n_{2k}}. 
\end{equation}
Let us further define $\mI_{\tau} = \{ i_{n_1}, i_{n_2},\ldots, i_{n_{2k}}\}$ and $\mI_{\tau}^c = \mI\setminus \mI_{\tau}$. To simplify the second and third set of deltas we recognize that in the definition of $\tilde{\delta}^{p}$ the symmetrization over the $i$ indices is unnecessary and the symmetrization over the $j$ can be restricted: splitting the action of $S_n$ on the $j$ indices as $S_n^{2k,n-2k}$ we can write the total contraction as
\begin{equation}
\label{eq:CCder1}
\begin{aligned}
C_{\mI_2 \mK_1}^{I_1}C_{\mK_1 \mI_1}^{I_2}  C_{\mJ_2 \mK_2}^{I_3}C_{\mK_2 \mJ_1}^{I_4}  
\sum_{k=0}^{\left \lfloor{\tfrac{n}{2}}\right \rfloor } \theta_{k} \sum_{\tau \in S_n^{\left(|\mI_1|,|\mI_2|\right)}(k)} \delta_{\tau} \frac{1}{n!} \times \\
\sum_{\substack{\sigma_1 \in S_{n-2k} \\ \rho \in S_n^{(2k,n-2k)}}} \delta^{(n-2k)}_{\mI^c_{\tau},\sigma_1\circ \rho\left(j_{2k+1}\right)\ldots \sigma_1\circ\rho\left(j_{n}\right)     }
\sum_{\sigma_2 \in S_{2k}} \delta_{\sigma_2\circ\rho\left(j_{1}\right) \sigma_2\circ\rho\left(j_{2}\right)}\ldots \delta_{\sigma_2\circ\rho\left(j_{2k-1}\right) \sigma_2\circ\rho\left(j_{2k}\right)},
\end{aligned}
\end{equation}
Note that the element $\rho$ effectively selects which $j$ indices occur in the delta functions together with the $i$'s. To simplify this expression further we need to identify which indices are contracted: this unfortunately complicates the expression even more, but we will see that the result is simple enough to work with. The idea is to split the elements $\sigma_1$ and $\sigma_2$ further into products using the decomposition of $S_{2k}$ and $S_{n-2k}$: depending on $\rho$ the $j$ indices are split into two parts containing the first $2k$ and last $n-2k$ elements, let us denote them as $\mJ_{2k}$ and $\mJ_{n-2k}$. The set $\mJ_{2k}$ describing the third set of delta functions overlaps with indices from $\mJ_1$ and $\mJ_2$, each of which forms a completely symmetric set of indices. Therefore symmetrizing indices belonging to $\mJ_1$ or $\mJ_2$ is unnecessary and we can restrict the sum over permutations to those which permute $\mJ_1$ indices to $\mJ_2$ indices and vice versa. A similar thing is true for the second set of delta functions. However, in this case there are two equivalent ways of doing it, although computationally usually one of the two ways is faster as it will result in less summands: as previously, the indices in $\mJ_{n-2k}$ overlap with $\mJ_1$ and $\mJ_2$, leading to a natural splitting of $S_{n-2k}$. However, the fact that in these delta functions the $j$ are always paired with an $i$ allows us to exploit the overlap of $\mI_{\tau}^c$ with $\mI_1$ and $\mI_2$ instead. Since this yields better results for the present case we use this second splitting. This leads to the following rewriting of \eqref{eq:CCder1}:
\begin{eqnarray}
& &C_{\mI_2 \mK_1}^{I_1}C_{\mK_1 \mI_1}^{I_2}  C_{\mJ_2 \mK_2}^{I_3}C_{\mK_2 \mJ_1}^{I_4} \sum_{k=0}^{\left \lfloor{\tfrac{n}{2}}\right \rfloor } \theta_{k} \sum_{\tau \in S_n^{\left(|\mI_1|,|\mI_2|\right)}(k)} \delta_{\tau} \frac{1}{n!} \times \nonumber \\ 
& &\sum_{\rho \in S_n^{(2k,n-2k)}} \sum_{\substack{\sigma_1^{(1)} \in S_{|\mI_1 \cap \mI_{\tau}^c|} \\ \sigma_1^{(2)} \in S_{|\mI_2 \cap \mI_{\tau}^c|} \\ \rho_1 \in S_{n-2k}^{(|\mI_1 \cap \mI_{\tau}^c|,|\mI_2 \cap \mI_{\tau}^c|)}}} \delta^{(n-2k)}_{\mI^c_{\tau}\,,\,\rho_1 \circ\sigma_1^{(1)}\circ\sigma_1^{(2)}\circ\rho\left(j_{2k+1}\right)\ldots \rho_1\circ \sigma_1^{(1)}\circ\sigma_1^{(2)}\circ\rho\left(j_{n}\right)     } \times \\
& & \sum_{\substack{\sigma_2^{(1)} \in S_{|\mJ_1|} \\ \sigma_2^{(2)} \in S_{|\mJ_2|} \\ \rho_2 \in S_{2k}^{(|\mJ_1 |,|\mJ_2| )}}} 
\delta_{\rho_2\circ \sigma_2^{(1)}\circ\sigma_2^{(2)}\circ\rho\left(j_{1}\right), \rho_2\circ \sigma_2^{(1)}\circ\sigma_2^{(2)}\circ\rho\left(j_{2}\right)}\ldots \delta_{\rho_2\circ \sigma_2^{(1)}\circ\sigma_2^{(2)}\circ\rho\left(j_{2k-1}\right), \rho_2\circ \sigma_2^{(1)}\circ\sigma_2^{(2)}\circ\rho\left(j_{2k}\right)}.\nonumber
\end{eqnarray}
Using the symmetry of the index sets $\mI_1$, $\mI_2$ and $\mJ_1$ and $\mJ_2$ we can now simplify this result, as the summands in the sums over the four different types of $\sigma$ permutations are all the same. This means that the following reduction is possible:
\begin{equation}
\begin{aligned}
 &C_{\mI_2 \mK_1}^{I_1}C_{\mK_1 \mI_1}^{I_2}  C_{\mJ_2 \mK_2}^{I_3}C_{\mK_2 \mJ_1}^{I_4} \sum_{k=0}^{\left \lfloor{\tfrac{n}{2}}\right \rfloor } \theta_{k} \sum_{\tau \in S_n^{\left(|\mI_1|,|\mI_2|\right)}(k)} \delta_{\tau} \frac{1}{n!}   \times  \\
&\sum_{\rho \in S_n^{(2k,n-2k)}} \sum_{\rho_1 \in S_{n-2k}^{(|\mI_1 \cap \mI_{\tau}^c|,|\mI_2 \cap \mI_{\tau}^c|)}} |\mI_1 \cap \mI_{\tau}^c|!\,|\mI_2 \cap \mI_{\tau}^c|!\, \delta^{(n-2k)}_{\mI^c_{\tau}\,,\,\rho_1 \circ\rho\left(j_{2k+1}\right)\ldots \rho_1\circ\rho\left(j_{n}\right)     } \times \\
& \sum_{ \rho_2 \in S_{2k}^{(|\mJ_1 |,|\mJ_2| )}} |\mJ_1|! \, |\mJ_2|!\,
\delta_{\rho_2\circ\rho\left(j_{1}\right), \rho_2\circ\rho\left(j_{2}\right)}\ldots \delta_{\rho_2\circ \rho\left(j_{2k-1}\right), \rho_2\circ \rho\left(j_{2k}\right)}
\end{aligned}
\end{equation}
and after noting that the final sum is still overcounting due to symmetries of the delta functions we note that finally we can rewrite our expression as
\begin{equation}
\begin{aligned}
\label{eq:CCsimplified}
 &C_{\mI_2 \mK_1}^{I_1}C_{\mK_1 \mI_1}^{I_2}  C_{\mJ_2 \mK_2}^{I_3}C_{\mK_2 \mJ_1}^{I_4} \sum_{k=0}^{\left \lfloor{\tfrac{n}{2}}\right \rfloor } \theta_{k} \sum_{\tau \in S_n^{\left(|\mI_1|,|\mI_2|\right)}(k)} \delta_{\tau} \frac{1}{n!} \times  \\
&\sum_{\rho \in S_n^{(2k,n-2k)}} \sum_{\rho_1 \in S_{n-2k}^{(|\mI_1 \cap \mI_{\tau}^c|,|\mI_2 \cap \mI_{\tau}^c|)}} |\mI_1 \cap \mI_{\tau}^c|!\,|\mI_2 \cap \mI_{\tau}^c|!\, \delta^{(n-2k)}_{\mI^c_{\tau}\,,\,\rho_1 \circ\rho\left(j_{2k+1}\right)\ldots \rho_1\circ\rho\left(j_{n}\right)     } \times \\
& \sum_{ \tau_2 \in S_{2k}^{(|\mJ_1 |,|\mJ_2| )}(k)\rho} |\mJ_1|! \, |\mJ_2|!\, 2^k k! \,
\delta_{\tau_2}.
\end{aligned}
\end{equation}
Here $\tau_2$ is an element of $S_{2k}^{(|\mJ_1 |,|\mJ_2| )}(k)\rho$, with which we mean that $\tau_2$ is one of the elements of $S_{2k}^{(|\mJ_1 |,|\mJ_2| )}(k)$ consisting of $k$ transpositions as it acts on the first $2k$ indices of the index set $\rho\left(\mJ\right)$. This is our final result, which plays an important role in the computation of the $\langle 3456\rangle$ correlator: as stated before, the computation of this tensor contraction was unfeasible with the previously known tools, rendering the computation of the $\langle 3456\rangle $ correlator from the Lagrangian infeasible as well. Using our new formula we were able to obtain the most complicated $a$ tensor for $\langle 3456\rangle$ with intermediate field with weight $9$ in $10$ minutes on a standard computer. 
\subsection{Reduction formulae}
\label{sec:reduction}
\subsubsection{Formulae}
There is no explicit summation formulae as eqn. \eqref{eq:B5Arutyunov} known for the product of $C$ tensors corresponding to $[1,k,1]$ and $[2,k,2]$ SO$(6)$ representations to simplify the computation of $t$ and $p$ tensors. However, one can generate the completeness conditions for sums involving lower weight representations. For the remaining sums one can obtain a system of linear equations, using the reduction formula \eqref{eq:redt}, \eqref{eq:redp} derived in \cite{Arutyunov:1999fb}: writing $f_i \equiv f\left(k_i\right) = k_i (k_i+4)$ they read\footnote{Summation over the fifth leg is assumed.} 
\bea\label{eq:redt}
	t_{125}t_{345}&=&-\frac{(f_1-f_2)(f_3-f_4)}{4f_5}a_{125}a_{345}-{1\ov 4}f_5(a_{135}a_{245}-a_{145}a_{235}),\nonumber\\
	f_5\,t_{125}t_{345}&=&-\frac{\left(f_1-f_2\right) \left(f_3-f_4\right) \left(f_5-3\right)}{4 f_5}a_{125} a_{345}\nonumber\\
	&& -\frac{1}{4}\left(f_1+f_2+f_3+f_4-f_5-3\right)f_5\left(a_{135} a_{245}-a_{145} a_{235}\right),\\
\nonumber\\	\label{eq:redp}
	p_{125}p_{345}&=& -\frac{\left(f_1-f_2\right) \left(f_3-f_4\right)}{2 \left(f_5-5\right)}t_{125} t_{345} -\frac{1}{20} \left(f_1+f_2-f_5\right) \left(f_3+f_4-f_5\right) a_{125} a_{345}\nonumber\\
	&& +\frac{1}{8}\left(f_1+f_3-f_5\right) \left(f_2+f_4-f_5\right)a_{135} a_{245}\nonumber\\
	&& +\frac{1}{8}\left(f_1+f_4-f_5\right)\left(f_2+f_3-f_5\right)a_{145} a_{235}-\frac{5}{4 \left(f_5-5\right) f_5}d_{125} d_{345} \nonumber\\
	f_5\,p_{125}p_{345}&=& -g_{1324}-g_{1423}-\frac{5\left(f_5-8\right)}{4 \left(f_5-5\right) f_5}d_{125} d_{345}-\frac{1}{2} \left(f_1-f_2\right) \left(f_3-f_4\right)t_{125} t_{345}\nonumber\\
	&& -\frac{1}{20}\left(f_1+f_2-f_5\right) \left(f_3+f_4-f_5\right) \left(f_5+2\right) a_{125} a_{345} \nonumber\\
	&& +\frac{1}{8}\left(f_1+f_2-6\right)\left(f_1+f_3-f_5\right)\left(f_2+f_4-f_5\right)a_{135} a_{245}\nonumber\\
	&& +\frac{1}{8}\left(f_1+f_2-6\right)\left(f_2+f_3-f_5\right)\left(f_1+f_4-f_5\right)a_{145} a_{235},
\eea
where we denoted
\bea
	g_{1234}&=&\frac{1}{4} \left(f_1+f_2-f_5-3\right) \left(f_3+f_4-f_5-3\right) t_{125} t_{345}\nonumber\\
	&&-\frac{ \left(f_1^2-\left(f_5-f_2\right){}^2\right) \left(\left(f_5-f_4\right){}^2-f_3^2\right)}{16 f_5}a_{125} a_{345},\nonumber\\
	d_{125}&=&\frac{1}{20}\left(-5 \left(f_1-f_2\right){}^2+3 f_5^2+2 \left(f_1+f_2\right) f_5\right)a_{125}.
\eea
We would also like to attract the reader's attention to the formula, derived in \cite{Arutyunov:2017dti}:
{\small
\bea\label{eq:redt2}
	&&(f_5-2)^2(t_{135}t_{245}+t_{145}t_{235})= \nonumber\\
 && -\frac{a_{135} a_{245} }{24
   f_5}\Big(-2 f_5^4+2 (3 f_1+3 f_2+3 f_3+3 f_4-28) f_5^3-2 (2
   f_1^2-(f_2-5 f_3-8 f_4+28) f_1 \nonumber \\
   &&+2 f_2^2+2 f_4^2+2 (f_3-12)
   (f_3-2)-(f_3+28) f_4+f_2 (8 f_3+5 f_4-28))
   f_5^2  \nonumber  \\
   &&+((f_2+3 f_3+4 f_4-12) f_1^2+(f_2^2+4 (f_3+f_4-20) f_2+3
   (f_3-4)^2+4 f_4^2+4 (f_3+4) f_4) f_1  \nonumber  \\
   &&+(3 f_2+f_3-12)
   f_4^2+4 ((f_3-3) f_2^2+(f_3 (f_3+4)+12) f_2-3
   (f_3-4) f_3)+(3 (f_2-4)^2 \nonumber  \\
   &&+f_3^2+4 (f_2-20)
   f_3) f_4) f_5+96 (f_1-f_3) (f_2-f_4)\Big)   \nonumber \\
   &&-\frac{a_{145} a_{235}}{24 f_5} \Big(-2 f_5^4+2 (3 f_1+3 f_2+3 f_3+3 f_4-28) f_5^3-2
   (2 f_1^2-(f_2-8 f_3-5 f_4+28) f_1 \nonumber \\
   &&+2 f_2^2+2 f_4^2+2 (f_3-12)
   (f_3-2)-(f_3+28) f_4+f_2 (5 f_3+8 f_4-28))
   f_5^2 \nonumber \\
   &&+((f_2+4 f_3+3 (f_4-4)) f_1^2+(f_2^2+4
   (f_3+f_4-20) f_2+4 f_3^2+3 (f_4-4)^2+4 f_3 (f_4+4))
   f_1 \nonumber \\
   &&+(4 f_2+f_3-12) f_4^2+3 (f_2-4) (f_3-4)
   (f_2+f_3) \nonumber \\
   &&+(4 f_2^2+4 (f_3+4) f_2+(f_3-80) f_3+48)
   f_4) f_5+96 (f_2-f_3) (f_1-f_4)\Big) \nonumber \\
  &&+\frac{  a_{125} a_{345}}{24}
  \Big(-4 f_5^3+4 (3 f_3+3 f_4-28) f_5^2-4 (2 f_4^2+5 f_3
   f_4-28 f_4+2 (f_3-14) f_3+48) f_5 \nonumber \\
   &&+6 (f_3-4) (f_4-4)
   (f_3+f_4)+f_2 (5 f_3^2+8 f_4 f_3-64 f_3+5 f_4^2+12 f_5^2-64 f_4 \nonumber \\
   &&-14
   (f_3+f_4-8) f_5+96)+f_1 (6 f_2^2+4 (2 (f_3+f_4-6)-5
   f_5) f_2+5 f_3^2+5 f_4^2+12 f_5^2 \nonumber \\
   &&-64 f_3+8 f_3 f_4-64 f_4-14 (f_3+f_4-8)
   f_5+96)+f_2^2 (5 f_3+5 f_4-8 (f_5+3)) \nonumber \\
   &&+f_1^2 (6 f_2+5 f_3+5
   f_4-8 (f_5+3))\Big)\, .
    \la{W}
 \eea
}
With its help, the \textit{sum} of two quadratic $tt$'s can be completely reduced to different products of $a$-tensors. 

In the current paper we obtain new reduction formulae for similar combinations of quadratic $pp$'s. They are a direct consequence of the following identities:
\bea\label{eq:redp2notcomplete}
	&&(f_5-2)^2\,p_{125}p_{345}=P_{1234}+R_{1234},\nonumber\\
	&&R_{1234}=-\frac{\left(f_1-f_2\right) \left(f_3-f_4\right) \left(f_5-7\right){}^2}{2 \left(f_5-5\right)}t_{125} t_{345} -\frac{\left(-5 \left(f_1-f_2\right){}^2+2 \left(f_1+f_2\right) f_5+3 f_5^2\right)}{320
   \left(f_5-5\right) f_5}\nonumber\\
   &&~~~~~~~~~~~~~~~~~\times \left(f_5-10\right){}^2 (-5 \left(f_3-f_4\right){}^2+3 f_5^2+2 \left(f_3+f_4\right) f_5)\, a_{125} a_{345},\nonumber\\
   &&P_{1234}=2 \left(Y_{1234}+Y_{1243}\right)-\frac{1}{20}\left(f_1+f_2-f_5\right) \left(f_3+f_4-f_5\right) f_5^2\, a_{125} a_{345}\nonumber\\
   &&~~~~~~~~~ -\frac{\left(f_2+f_3-f_5\right) \left(f_1+f_4-f_5\right)}{16 f_5} \Big(\left(f_1+f_2+f_3+f_4-16\right) f_5^2 \\
   &&~~~~~~~~~ -(f_1^2+2 \left(f_2-8\right) f_1+\left(f_2-16\right) f_2+\left(f_3+f_4-16\right)
   \left(f_3+f_4\right)) f_5\nonumber\\
   &&~~~~~~~~~ -128 f_5+\left(f_2-f_3\right) \left(f_1-f_4\right) \left(f_1+f_2+f_3+f_4-16\right)\Big)\, a_{145} a_{235}\nonumber\\
   &&~~~~~~~~~ -\frac{ \left(f_1+f_3-f_5\right) \left(f_2+f_4-f_5\right)}{16 f_5} \Big(\left(f_1+f_2+f_3+f_4-16\right) f_5^2 \nonumber\\
   &&~~~~~~~~~ -\left(f_1^2+2 \left(f_2-8\right) f_1+\left(f_2-16\right) f_2+\left(f_3+f_4-16\right) \left(f_3+f_4\right)\right) f_5\nonumber\\
   &&~~~~~~~~~ -128 f_5+\left(f_1-f_3\right) \left(f_2-f_4\right)
   \left(f_1+f_2+f_3+f_4-16\right) \Big)a_{135} a_{245}\nonumber\\
   &&~~~~~~~~~ -\frac{1}{4} \left(f_1+f_2+f_3+f_4-16\right) \left(f_1+f_3-f_5-3\right)\left(f_2+f_4-f_5-3\right) t_{135} t_{245}\nonumber\\
   &&~~~~~~~~~ -\frac{1}{4} \left(f_1+f_2+f_3+f_4-16\right) \left(f_2+f_3-f_5-3\right) \left(f_1+f_4-f_5-3\right) t_{145} t_{235},\nonumber
\eea
where $Y_{1234}$ is defined by \eqref{eq:Y} and has the following symmetry properties:
\bea
	 Y_{1234}=Y_{1324}=Y_{2143}=Y_{3412}.
\eea
 We were not able to reduce $Y_{1234}$, but rather $Y_{12[34]}\equiv\dfrac{1}{2}\left(Y_{1234}-Y_{1243}\right)$:
\bea\label{eq:Ya}
	Y_{12[34]}&=&\frac{1}{8} \left(f_1+f_2-f_5-5\right) \left(f_3+f_4-f_5-3\right) \left(f_5+3\right)\,t_{125}t_{345}\nonumber\\
	&& -\frac{1}{8} \left(f_1+f_3-f_5-3\right) \left(f_2+f_4-f_5-3\right) t_{135} t_{245}\nonumber\\
	&& +\frac{1}{8} \left(f_2+f_3-f_5-3\right) \left(f_1+f_4-f_5-3\right) t_{145} t_{235}\nonumber\\
	&& \frac{1}{32 f_5} \left(f_1^2-\left(f_3-f_5\right){}^2\right) \left(\left(f_4-f_5\right){}^2-f_2^2\right)a_{135} a_{245}\nonumber\\
	&&-\frac{1}{32 f_5}\left(f_2^2-\left(f_3-f_5\right){}^2\right)
   \left(\left(f_4-f_5\right){}^2-f_1^2\right)a_{145} a_{235}.
\eea
Thus, using \eqref{eq:redp2notcomplete}-\eqref{eq:Ya}, one is able, similar to \eqref{eq:redt2}, to find an expression only for the \textit{difference} of two quadratic $pp$'s. In fact, it has the following structure (schematically):
\bea\label{eq:redp2}
	f_5^2(p_{135}p_{245}-p_{145}p_{235})= {1\ov 2}f_5^3\,t_{125}t_{345}+\{...\},
\eea
where \{...\} contains only quadratic $tt$'s and different $aa$'s.

From \eqref{eq:redp2} it directly follows another useful reduction formula:
\bea
	 &&f_5^3(t_{1234}-t_{1324}+t_{1423})=\Sigma_{1234}-\Sigma_{1324}+\Sigma_{1423},\nonumber\\
	 && \Sigma_{1234} = (f_1+f_2+f_3+f_4-13)f_5^2\,t_{125}t_{345}\nonumber\\
	 &&~~~~~~~~~ +\Big((10 f_3+10 f_4-f_1 (f_3+f_4-8)-f_2(f_3+f_4-8)-51) f_5\nonumber\\
	 &&~~~~~~~~~~~~~~~~ +(-5 f_1-5 f_2+21)(f_3+f_4-3)\Big)t_{125} t_{345}\\
	 &&~~~~~~~~~ +\frac{\left(f_3-f_4\right) \left(\left(f_2-f_5\right){}^2-f_1^2\right) \left(f_3+f_4-f_5\right)}{2 f_5}a_{125} a_{345}.\nonumber
\eea
\subsubsection{Proof}
Let us show how the new formula \eqref{eq:redp2notcomplete} can be obtained\footnote{We thank Sergey Frolov for sharing this calculation with us.}. We closely follow \cite{Arutyunov:1999fb} and \cite{Arutyunov:2017dti}.
To do the calculation we need to use 
\bal
c_{125}&=\int \na^{\a}\na^{\b}Y^{I_1}\na_{\a}\na_{\b}Y^{I_2}Y^{I_5}= \frac{1}{4} \left(f_1+f_2-f_5-8\right) \left(f_1+f_2-f_5\right)a_{125},\\
&\mu_{125}=\frac{f_2-f_1}{f_5-5}t_{125},
\quad \nu_{125}=\frac{5}{4f_5(f_5-5)}  d_{125},\\
& d_{125}= \frac{1}{20}\left(-5 \left(f_1-f_2\right){}^2+3 f_5^2+2 \left(f_1+f_2\right) f_5\right)a_{125}.
\eal
which are used in the relation
\bal
&\na_{(\a} Y^1\na_{\b )} Y^2=p_{125}Y_{(\a\b )}^5+
\mu_{125}\na_{(\a }Y_{\b )}^5 +\nu_{125}\na_{(\a }\na_{\b )}Y^5 \,.
\la{tendec},
\eal
where from now on the round bracket denote \emph{traceless} symmetrization. Then we need the formulae
\bal
	& \na^2 Y^5 = -f_5 Y^5\,,\quad 
\na^2 \na_\a Y^5 = (4-f_5)  \na_\a Y^5\,,\\
	& \na^2  \na_\a \na_\b Y^5 = (10-f_5) \na_\a \na_\b Y^5+2 g_{\a\b}f_5Y^5\,, \\
	& \na^2  \na_{(\a} \na_{\b)} Y^5 = (10-f_5)  \na_{(\a} \na_{\b)}  Y^5\,,\\
	& \na^2 Y_\a^5 = (1-f_5) Y_\a^5\,,\quad \na^2  \na_\a Y_\b^5 = (5-f_5)  \na_\a Y_\b^5+2 \na_\b Y_\a^5\,,\\
	& \na^2  \na_{(\a} Y_{\b)}^5 = (7-f_5)  \na_{(\a} Y_{\b)}^5\,, \quad \na^2 Y_{(\a\b)}^5 = (2-f_5) Y_{(\a\b)}^5\,.
\eal
These formulae and \eqref{tendec} give
\bal
& (2-f_5)p_{125}Y_{(\a\b )}^5=  \na^2\big(\na_{(\a} Y^1\na_{\b )} Y^2\big)-
(7-f_5)\mu_{125}\na_{(\a }Y_{\b )}^5 -(10-f_5)\nu_{125}\na_{(\a }\na_{\b )}Y^5 \,,
\la{tendec2}
\eal
and
\bal
(&2-f_5)^2p_{125}p_{345} = \int  \na^2\big(\na_{(\a} Y^1\na_{\b )} Y^2\big) \na^2\big(\na_{(\a} Y^3\na_{\b )} Y^4\big)\\
&-(7-f_5)^2\mu_{125}\int   \na_{(\a }Y_{\b )}^5\na_{(\a} Y^3\na_{\b )} Y^4-(7-f_5)^2\mu_{345}\int   \na_{(\a }Y_{\b )}^5\na_{(\a} Y^1\na_{\b )} Y^2
\\
&-(10-f_5)^2\nu_{125}\int  \na_{(\a }\na_{\b )}Y^5\na_{(\a} Y^3\na_{\b )} Y^4-(10-f_5)^2\nu_{345}\int  \na_{(\a }\na_{\b )}Y^5\na_{(\a} Y^1\na_{\b )} Y^2\\
&-{1\ov2}(5-f_5)(7-f_5)^2\mu_{125}\mu_{345}-{4\ov5}f_5(5-f_5)(10-f_5)^2\nu_{125}\nu_{345}\,,
\eal
where we took into account that
\bal
&\int   \na_{(\a }Y_{\b )}^5 \na_{(\a }Y_{\b )}^5 = -{1\ov2}(5-f_5)\,,\\
&\int  \na_{(\a }\na_{\b )}Y^5\na_{(\a }\na_{\b )}Y^5=-{4\ov5}f_5(5-f_5)\,,\\
&\int   \na_{(\a }Y_{\b )}^5\na_{(\a }\na_{\b )}Y^5 =0\,.
\eal
Using integration by parts and the definitions \eqref{eq:aptintegrals} and 
\begin{equation}
b_{123} = \int \nabla^{\a} Y^{I_1} \nabla_{\a} Y^{I_2} Y^{I_3} = \tfrac{1}{2} \left( f_1 +f_2 -f_3\right) a_{123} 
\end{equation}
we can reduce the second and the third line to $t$ and $a$ contributions:
\bal
	(2-f_5)^2p_{125}p_{345}& = \int  \na^2\big(\na_{(\a} Y^1\na_{\b )} Y^2\big) \na^2\big(\na_{(\a} Y^3\na_{\b )} Y^4\big) + R_{1234},\\
	R_{1234}&=-{1\ov2}(10-f_5)^2\nu_{125}(f_3b_{453}+f_4b_{354}-f_5b_{345})\\
	& -{1\ov2}(10-f_5)^2\nu_{345}(f_1b_{251}+f_2b_{152}-f_5b_{125})\\
	& +{1\ov2}(5-f_5)(7-f_5)^2\mu_{125}\mu_{345}-{4\ov5}f_5(5-f_5)(10-f_5)^2\nu_{125}\nu_{345}\\
	& -{f_5\ov 5}(10-f_5)^2(b_{125}\nu_{345}+\nu_{125}b_{345})\,.
\eal

So, the main problem is to evaluate
\bal
P_{1234}\equiv \int  \na^2\big(\na_{(\a} Y^1\na_{\b )} Y^2\big) \na^2\big(\na_{(\a} Y^3\na_{\b )} Y^4\big). 
\eal
We have
\bal
	P_{1234}&={1\ov2} \int  \na^2\big(\na_{\a} Y^1\na_{\b } Y^2\big) \na^2\big(\na_{\a} Y^{3}\na_{\b} Y^{4}+\na_{\a} Y^{4}\na_{\b} Y^{3}-{2\ov5}g_{\a\b}\na_{\g} Y^{3}\na_{\g} Y^{4}\big) \\
&={1\ov2} \int \na^2\big(\na_{\a} Y^1\na_{\b } Y^2\big) \na^2\big(\na_{\a} Y^{3}\na_{\b} Y^{4}+\na_{\a} Y^{4}\na_{\b} Y^{3}\big)\\
	&~~~ -{1\ov5} \int \na^2\big(\na_{\a} Y^1\na_{\a } Y^2\big)\na^2\big(\na_{\g} Y^{3}\na_{\g} Y^{4}\big)\\
	&={1\ov2} \int  \na^2\big(\na_{\a} Y^1\na_{\b } Y^2\big) \na^2\big(\na_{\a} Y^{3}\na_{\b} Y^{4}+\na_{\a} Y^{4}\na_{\b} Y^{3}\big)-{1\ov5}f_5^2b_{125}b_{345}\,,\\
P_{1234}&={1\ov2} \int\Big((8-f_1-f_2)\na_{\a} Y^1\na_{\b } Y^2 + 2\na_{\g}\na_{\a} Y^1\na_{\g}\na_{\b } Y^2\Big)\\
	&\quad\qquad\times  \Big((8-f_3-f_4)\na_{\a} Y^3\na_{\b } Y^4 + 2\na_{\r}\na_{\a} Y^3\na_{\r}\na_{\b } Y^4\\
	&\qquad\qquad +(8-f_3-f_4)\na_{\a} Y^4\na_{\b } Y^3 + 2\na_{\r}\na_{\a} Y^4\na_{\r}\na_{\b } Y^3\Big)-{1\ov5}f_5^2b_{125}b_{345},\\
	P_{1234}&=P_{1234}^{(8)}+P_{1234}^{(6)}+P_{1234}^{(4)}-{1\ov5}f_5^2b_{125}b_{345} \,,
\eal
where we denote
\bal
	P_{1234}^{(4)}&\equiv {1\ov2} \int(8-f_1-f_2)(8-f_3-f_4)\na_{\a} Y^1\na_{\b } Y^2  \big(\na_{\a} Y^3\na_{\b } Y^4 +\na_{\a} Y^4\na_{\b } Y^3\big)\,,\\
	P_{1234}^{(6)}&\equiv \int \Big( (8-f_1-f_2)\big(\na_{\a} Y^1\na_{\b } Y^2\na_{\r}\na_{\a} Y^3\na_{\r}\na_{\b } Y^4 + \na_{\a} Y^1\na_{\b } Y^2\na_{\r}\na_{\a} Y^4\na_{\r}\na_{\b } Y^3\big)\\
	&+(8-f_3-f_4)\big(\na_{\a} Y^3\na_{\b } Y^4\na_{\g}\na_{\a} Y^1\na_{\g}\na_{\b } Y^2 + \na_{\a} Y^4\na_{\b } Y^3\na_{\g}\na_{\a} Y^1\na_{\g}\na_{\b } Y^2   \big)\Big) \,,\\
	P_{1234}^{(8)}&\equiv 2 \int \na_{\g}\na_{\a} Y^1\na_{\g}\na_{\b } Y^2\big(\na_{\r}\na_{\a} Y^3\na_{\r}\na_{\b } Y^4+\na_{\r}\na_{\a} Y^4\na_{\r}\na_{\b } Y^3\big) \,.	
\eal
We immediately find 
\bal
	P_{1234}^{(4)}= {1\ov2}(8-f_1-f_3)(8-f_2-f_4)(b_{135}b_{245}+b_{145}b_{235}) \,.
\eal
To reduce the six-derivative terms we use
\bal
	&\na_\b\na_\a Y^1\na_\b Y^2 = t_{125}^{(2)}\,Y^5_\a + b_{152}^{(2)}\,\na_\a Y^5\,,\\
	&t_{125}^{(2)}={1\ov2}(f_1+f_2-f_5-3)t_{125}\,,\\
	&b_{152}^{(2)}={(f_1+f_2-f_5)(f_1-f_2+f_5)\ov 4f_5}a_{125}\,.
\eal
Then we get
\bal
	P_{1234}^{(6)}&= (8-f_1-f_2)\big(t_{135}^{(2)}t_{245}^{(2)} +f_5b_{351}^{(2)}b_{452}^{(2)} + t_{145}^{(2)}t_{235}^{(2)} +f_5b_{451}^{(2)}b_{352}^{(2)} \big)\\
	&+(8-f_3-f_4)\big(t_{135}^{(2)}t_{245}^{(2)} +f_5b_{153}^{(2)}b_{254}^{(2)}+t_{145}^{(2)}t_{235}^{(2)} +f_5b_{154}^{(2)}b_{253}^{(2)} \big).
\eal
Introducing
\bal\label{eq:Y}
Y_{1234}&=\int \na_{\g}\na_{\a} Y^1\na_{\g}\na_{\b } Y^2\na_{\r}\na_{\a} Y^3\na_{\r}\na_{\b } Y^4\,,
\eal
we find the last term
\bal
	P_{1234}^{(8)}&=2(Y_{1234}+Y_{1243})\,.
\eal

It remains to compute $Y_{12[34]}\equiv\dfrac{1}{2}\left(Y_{1234}-Y_{1243}\right)$. Integration by parts gives:
\bal\label{eq:Yacomp}
	Y_{12[34]}=Y_{[12]34}=&\na_{\a}\na_{\b} Y^{[1} \na_{\g}\na^{\b} Y^{2]}\na^{\a}\na_{\de} Y^{3} \na^{\g}\na^{\de} Y^{4}\\
	=&-\na_{\g}\na_{\b}\na_{\a} Y^{[1} \na^{\g}\na^{\b} Y^{2]}\na^{\a}\na_{\de} Y^{3} \na^{\de} Y^{4}\\
	&-\na_{\a}\na_{\b} Y^{[1} \na^{2}\na^{\b} Y^{2]}\na^{\a}\na_{\de} Y^{3} \na^{\de} Y^{4}\\
	&-\na_{\a}\na_{\b} Y^{[1} \na_{\g}\na^{\b} Y^{2]}\na^{\g}\na^{\a}\na_{\de} Y^{3} \na^{\de} Y^{4}\,.
\eal
With the help of the following identity:
\bea
	\left[\na_\a\,,\, \na_\b\right] \xi_\g=g_{\a\g}\xi_\b-g_{\b\g}\xi_\a\,,
\eea 
one finds the last line in \eqref{eq:Yacomp}:
\bal
	\na_{\a}\na_{\b} Y^{[1}& \na_{\g}\na^{\b} Y^{2]}\na^{\g}\na^{\a}\na_{\de} Y^{3} \na^{\de} Y^{4}={1\ov 2}\na_{\a}\na_{\b} Y^{[1} \na_{\g}\na^{\b} Y^{2]}[\na^{\g},\na^{\a}]\na_{\de} Y^{3} \na^{\de} Y^{4}\nonumber\\
	&=\na_{\a}\na_{\b} Y^{[1} \na_{\g}\na^{\b} Y^{2]}\na^{\a} Y^{[3} \na^{\g} Y^{4]}\,.
\eal
Finally, using
\bal
	&\na_\g\na_\b\na_\a Y^1\na_\g\na_\b Y^2 =t_{125}^{(4)}\,Y^5_\a + b_{152}^{(4)}\,\na_\a Y^5\,,\\
	&t_{125}^{(4)}={1\ov4}(f_1+f_2-f_5-3)(f_1+f_2-f_5-13)t_{125} + f_1t_{125}\,,\\
	&b_{152}^{(4)}={(f_1+f_2-f_5)(f_1+f_2-f_5-10)\ov 4f_5}b_{152}-{f_1\ov f_5}b_{251}\,.
\eal
we find 
\bal
	Y_{12[34]}&=-{1\ov2}(t^{(4)}_{125}-t^{(4)}_{215})t^{(2)}_{345}-{1\ov2}f_5 (b^{(4)}_{152}-b^{(4)}_{251})b^{(2)}_{354}\\
	& -{1\ov2}\big( (4-f_2)t^{(2)}_{125}-(4-f_1)t^{(2)}_{215} \big)t^{(2)}_{345}-{1\ov2}f_5 \big( (4-f_2)b^{(2)}_{152}-(4-f_1)b^{(2)}_{251} \big)b^{(2)}_{354}\\
	& -{1\ov2}(t^{(2)}_{135}t^{(2)}_{245}-t^{(2)}_{145}t^{(2)}_{235})-{1\ov 2}f_5(b^{(2)}_{153}b^{(2)}_{254}-b^{(2)}_{154}b^{(2)}_{253})\,.
\eal

We would also like to note, that the method, similar to one in \citep{Arutyunov:2017dti}, gives exactly the same relations between cubic $tt$'s, as those derived from the relations for quadratic $pp$'s. 

\section{Intermediate results}\label{app:intermediate results}
\subsection{$\langle CCC\rangle\langle CCC\rangle$-products}
We give explicit results for the products of $C$ tensors. 
\paragraph{$\langle 2345\rangle$ weights}
\bal
	&\langle C_2^1C_3^2C_{[0,1,0]}^5\rangle\langle C_4^3C_5^4C_{[0,1,0]}^5\rangle=T_6\,; \\
	&\langle C_2^1C_3^2C_{[0,3,0]}^5\rangle\langle C_4^3C_5^4C_{[0,3,0]}^5\rangle=\frac{2 T_4}{3}+\frac{T_5}{3}-\frac{T_6}{6}\, ;\\
	&\langle C_2^1C_3^2C_{[0,5,0]}^5\rangle\langle C_4^3C_5^4C_{[0,5,0]}^5\rangle=\frac{3 T_1}{10}+\frac{3 T_2}{5}+\frac{T_3}{10}-\frac{T_4}{5}-\frac{T_5}{10}+\frac{T_6}{50};
\eal

\bal
	&\langle C_2^1C_4^3C_{[0,2,0]}^5\rangle\langle C_3^2C_5^4C_{[0,2,0]}^5\rangle= T_3\, ;\\
	&\langle C_2^1C_4^3C_{[0,4,0]}^5\rangle\langle C_3^2C_5^4C_{[0,4,0]}^5\rangle= \frac{3 T_2}{4}-\frac{3 T_3}{20}+\frac{T_5}{4}\, ;\\
	&\langle C_2^1C_4^3C_{[0,6,0]}^5\rangle\langle C_3^2C_5^4C_{[0,6,0]}^5\rangle= \frac{2 T_1}{5}-\frac{8 T_2}{35}+\frac{2 T_3}{105}+\frac{8 T_4}{15}-\frac{8 T_5}{105}+\frac{T_6}{15};
\eal

\bal
	&\langle C_2^1C_5^4C_{[0,3,0]}^5\rangle\langle C_3^2C_4^3C_{[0,3,0]}^5\rangle= T_1\,; \\
	&\langle C_2^1C_5^4C_{[0,5,0]}^5\rangle\langle C_3^2C_4^3C_{[0,5,0]}^5\rangle= -\frac{T_1}{5}+\frac{3 T_2}{5}+\frac{2 T_4}{5}\,; \\
	&\langle C_2^1C_5^4C_{[0,7,0]}^5\rangle\langle C_3^2C_4^3C_{[0,7,0]}^5\rangle= \frac{3 T_1}{98}-\frac{3 T_2}{14}+\frac{2 T_3}{7}-\frac{T_4}{7}+\frac{4 T_5}{7}+\frac{T_6}{7};
\eal

\bal
	&\langle C_2^1C_3^2C_{[1,1,1]}^5\rangle\langle C_4^3C_5^4C_{[1,1,1]}^5\rangle= -\frac{3 T_4}{2}+\frac{3 T_5}{2}-\frac{3 T_6}{10}\,; \\
	&\langle C_2^1C_3^2C_{[1,3,1]}^5\rangle\langle C_4^3C_5^4C_{[1,3,1]}^5\rangle=-\frac{5 T_1}{6}+\frac{5 T_2}{12}+\frac{5 T_3}{12}+\frac{5 T_4}{21}-\frac{25 T_5}{84}+\frac{T_6}{42} ;
\eal

\bal
	&\langle C_2^1C_4^3C_{[1,2,1]}^5\rangle\langle C_3^2C_5^4C_{[1,2,1]}^5\rangle= -\frac{4 T_2}{3}-\frac{4 T_3}{9}+\frac{4 T_5}{3} \,; \\
	&\langle C_2^1C_4^3C_{[1,4,1]}^5\rangle\langle C_3^2C_5^4C_{[1,4,1]}^5\rangle=  -\frac{9 T_1}{10}+\frac{9 T_2}{80}+\frac{3 T_3}{80}+\frac{3 T_4}{5}-\frac{21 T_5}{80}+\frac{3 T_6}{10};
\eal

\bal
	&\langle C_2^1C_5^4C_{[1,3,1]}^5\rangle\langle C_3^2C_4^3C_{[1,3,1]}^5\rangle= -\frac{5 T_1}{28}-\frac{5 T_2}{4}+\frac{5 T_4}{4} \,; \\
	&\langle C_2^1C_5^4C_{[1,5,1]}^5\rangle\langle C_3^2C_4^3C_{[1,5,1]}^5\rangle= \frac{T_1}{45}+\frac{7 T_2}{30}-\frac{7 T_3}{10}-\frac{14 T_4}{45}+\frac{7 T_5}{30}+\frac{7 T_6}{15};
\eal

\bal
	&\langle C_2^1C_3^2C_{[2,1,2]}^5\rangle\langle C_4^3C_5^4C_{[2,1,2]}^5\rangle= \frac{8 T_1}{9}-\frac{16 T_2}{9}+\frac{8 T_3}{9}-\frac{16 T_4}{63}-\frac{40 T_5}{63}+\frac{8 T_6}{63}\,;
\eal

\bal
	&\langle C_2^1C_4^3C_{[2,2,2]}^5\rangle\langle C_3^2C_5^4C_{[2,2,2]}^5\rangle= \frac{3 T_1}{5}-\frac{3 T_2}{40}+\frac{33 T_3}{280}-\frac{6 T_4}{5}-\frac{21 T_5}{40}+\frac{3 T_6}{5} \,;
\eal

\bal
	&\langle C_2^1C_5^4C_{[2,3,2]}^5\rangle\langle C_3^2C_4^3C_{[2,3,2]}^5\rangle= \frac{8 T_1}{75}-\frac{16 T_2}{75}+\frac{16 T_3}{25}-\frac{32 T_4}{75}-\frac{32 T_5}{25}+\frac{16 T_6}{25} \,.
\eal
\paragraph{$\langle 3456 \rangle$ weights}
\bal
	\langle C_3^1C_4^2C_{[0,1,0]}^5\rangle\langle C_5^3C_6^4C_{[0,1,0]}^5\rangle &= T_{10} \,; \\
	\langle C_3^1C_4^2C_{[0,3,0]}^5\rangle\langle C_5^3C_6^4C_{[0,3,0]}^5\rangle &= \frac{2 T_8}{3}+\frac{T_9}{3}-\frac{T_{10}}{6} \,;\\
	\langle C_3^1C_4^2C_{[0,5,0]}^5\rangle\langle C_5^3C_6^4C_{[0,5,0]}^5\rangle &= \frac{3 T_5}{10}+\frac{3 T_6}{5}+\frac{T_7}{10}-\frac{T_8}{5}-\frac{T_9}{10}+\frac{T_{10}}{50} \,;\\
	\langle C_3^1C_4^2C_{[0,7,0]}^5\rangle\langle C_5^3C_6^4C_{[0,7,0]}^5\rangle &= \frac{4 T_1}{35}+\frac{18 T_2}{35}+\frac{12 T_3}{35}+\frac{T_4}{35}-\frac{9 T_5}{70}-\frac{9 T_6}{35}-\frac{3 T_7}{70}\\
	&+\frac{9 T_8}{245}+\frac{9 T_9}{490}-\frac{T_{10}}{490} \,;
\eal

\bal
	\langle C_3^1C_5^3C_{[0,2,0]}^5\rangle\langle C_4^2C_6^4C_{[0,2,0]}^5\rangle &= T_4 \, ;\\
	\langle C_3^1C_5^3C_{[0,4,0]}^5\rangle\langle C_4^2C_6^4C_{[0,4,0]}^5\rangle &= \frac{3 T_3}{4}-\frac{3 T_4}{20}+\frac{T_7}{4} \, ;\\
	\langle C_3^1C_5^3C_{[0,6,0]}^5\rangle\langle C_4^2C_6^4C_{[0,6,0]}^5\rangle &= \frac{2 T_2}{5}-\frac{8 T_3}{35}+\frac{2 T_4}{105}+\frac{8 T_6}{15}-\frac{8 T_7}{105}+\frac{T_9}{15} \,;\\
	\langle C_3^1C_5^3C_{[0,8,0]}^5\rangle\langle C_4^2C_6^4C_{[0,8,0]}^5\rangle &= \frac{5 T_1}{28}-\frac{5 T_2}{28}+\frac{5 T_3}{112}-\frac{5 T_4}{2352}+\frac{15 T_5}{28}-\frac{5 T_6}{21}+\frac{5 T_7}{336} \\
	& +\frac{15 T_8}{56}-\frac{5 T_9}{168}+\frac{T_{10}}{56} \,;\\
\eal

\bal
	\langle C_3^1C_6^4C_{[0,3,0]}^5\rangle\langle C_4^2C_5^3C_{[0,3,0]}^5\rangle &=  T_1 \,; \\
	\langle C_3^1C_6^4C_{[0,5,0]}^5\rangle\langle C_4^2C_5^3C_{[0,5,0]}^5\rangle &= -\frac{T_1}{5}+\frac{3 T_2}{5}+\frac{2 T_5}{5} \,; \\
	\langle C_3^1C_6^4C_{[0,7,0]}^5\rangle\langle C_4^2C_5^3C_{[0,7,0]}^5\rangle &= \frac{3 T_1}{98}-\frac{3 T_2}{14}+\frac{2 T_3}{7}-\frac{T_5}{7}+\frac{4 T_6}{7}+\frac{T_8}{7} \,;\\
	\langle C_3^1C_6^4C_{[0,9,0]}^5\rangle\langle C_4^2C_5^3C_{[0,9,0]}^5\rangle &= -\frac{T_1}{252}+\frac{T_2}{21}-\frac{T_3}{7}+\frac{5 T_4}{42}+\frac{2 T_5}{63}-\frac{2 T_6}{7}+\frac{10 T_7}{21}\\
	&-\frac{T_8}{14}+\frac{5 T_9}{14}+\frac{T_{10}}{21} \,;
\eal

\bal
	\langle C_3^1C_4^2C_{[1,1,1]}^5\rangle\langle C_5^3C_6^4C_{[1,1,1]}^5\rangle &= -\frac{3 T_8}{2}+\frac{3 T_9}{2}-\frac{3 T_{10}}{10} \,; \\
	\langle C_3^1C_4^2C_{[1,3,1]}^5\rangle\langle C_5^3C_6^4C_{[1,3,1]}^5\rangle &= -\frac{5 T_5}{6}+\frac{5 T_6}{12}+\frac{5 T_7}{12}+\frac{5 T_8}{21}-\frac{25 T_9}{84}+\frac{T_{10}}{42} \,;\\
	\langle C_3^1C_4^2C_{[1,5,1]}^5\rangle\langle C_5^3C_6^4C_{[1,5,1]}^5\rangle &= -\frac{7 T_1}{20}-\frac{7 T_2}{20}+\frac{7 T_3}{12}+\frac{7 T_4}{60}+\frac{49 T_5}{180}-\frac{7 T_6}{45}-\frac{77 T_7}{540}\\
	&-\frac{17 T_8}{540}+\frac{23 T_9}{540}-\frac{T_{10}}{540} \,;
\eal

\bal
	\langle C_3^1C_5^3C_{[1,2,1]}^5\rangle\langle C_4^2C_6^4C_{[1,2,1]}^5\rangle &= -\frac{4 T_3}{3}-\frac{4 T_4}{9}+\frac{4 T_7}{3} \, ;\\
	\langle C_3^1C_5^3C_{[1,4,1]}^5\rangle\langle C_4^2C_6^4C_{[1,4,1]}^5\rangle &= -\frac{9 T_2}{10}+\frac{9 T_3}{80}+\frac{3 T_4}{80}+\frac{3 T_6}{5}-\frac{21 T_7}{80}+\frac{3 T_9}{10} \, ;\\
	\langle C_3^1C_5^3C_{[1,6,1]}^5\rangle\langle C_4^2C_6^4C_{[1,6,1]}^5\rangle &= -\frac{16 T_1}{35}+\frac{48 T_2}{175}-\frac{12 T_3}{1225}-\frac{4 T_4}{1225}-\frac{16 T_5}{105}-\frac{128 T_6}{525}+\frac{148 T_7}{3675}\\
	&+\frac{8 T_8}{15}-\frac{8 T_9}{75}+\frac{8 T_{10}}{105} \,;
\eal

\bal
	\langle C_3^1C_6^4C_{[1,3,1]}^5\rangle\langle C_4^2C_5^3C_{[1,3,1]}^5\rangle &= -\frac{5 T_1}{28}-\frac{5 T_2}{4}+\frac{5 T_5}{4} \,; \\
	\langle C_3^1C_6^4C_{[1,5,1]}^5\rangle\langle C_4^2C_5^3C_{[1,5,1]}^5\rangle &= \frac{T_1}{45}+\frac{7 T_2}{30}-\frac{7 T_3}{10}-\frac{14 T_5}{45}+\frac{7 T_6}{30}+\frac{7 T_8}{15} \,; \\
	\langle C_3^1C_6^4C_{[1,7,1]}^5\rangle\langle C_4^2C_5^3C_{[1,7,1]}^5\rangle &= -\frac{3 T_1}{1232}-\frac{45 T_2}{1232}+\frac{81 T_3}{308}-\frac{9 T_4}{28}+\frac{69 T_5}{1232}-\frac{9 T_6}{77}-\frac{9 T_7}{28}\\
	&-\frac{117 T_8}{616}+\frac{27 T_9}{56}+\frac{9 T_{10}}{56} \,;
\eal

\bal
	\langle C_3^1C_4^2C_{[2,1,2]}^5\rangle\langle C_5^3C_6^4C_{[2,1,2]}^5\rangle &= \frac{8 T_5}{9}-\frac{16 T_6}{9}+\frac{8 T_7}{9}-\frac{16 T_8}{63}-\frac{40 T_9}{63}+\frac{8 T_{10}}{63} \,; \\
	\langle C_3^1C_4^2C_{[2,3,2]}^5\rangle\langle C_5^3C_6^4C_{[2,3,2]}^5\rangle &= \frac{16 T_1}{25}-\frac{24 T_2}{25}+\frac{8 T_4}{25}-\frac{112 T_5}{225}-\frac{16 T_6}{225}-\frac{88 T_7}{225}\\
	&+\frac{32 T_8}{225}+\frac{32 T_9}{225}-\frac{4 T_{10}}{225} \,;
\eal

\bal
	\langle C_3^1C_5^3C_{[2,2,2]}^5\rangle\langle C_4^2C_6^4C_{[2,2,2]}^5\rangle &= \frac{3 T_2}{5}-\frac{3 T_3}{40}+\frac{33 T_4}{280}-\frac{6 T_6}{5}-\frac{21 T_7}{40}+\frac{3 T_9}{5} \, ;\\
	\langle C_3^1C_5^3C_{[2,4,2]}^5\rangle\langle C_4^2C_6^4C_{[2,4,2]}^5\rangle &= \frac{64 T_1}{105}-\frac{64 T_2}{175}+\frac{64 T_3}{525}-\frac{64 T_4}{3675}-\frac{64 T_5}{63}-\frac{256 T_6}{1575}+\frac{64 T_7}{525}\\
	&+\frac{64 T_8}{315}-\frac{64 T_9}{225}+\frac{64 T_{10}}{315} \, ;
\eal

\bal
	\langle C_3^1C_6^4C_{[2,3,2]}^5\rangle\langle C_4^2C_5^3C_{[2,3,2]}^5\rangle &= \frac{8 T_1}{75}-\frac{16 T_2}{75}+\frac{16 T_3}{25}-\frac{32 T_5}{75}-\frac{32 T_6}{25}+\frac{16 T_8}{25} \,; \\
	\langle C_3^1C_6^4C_{[2,5,2]}^5\rangle\langle C_4^2C_5^3C_{[2,5,2]}^5\rangle &= -\frac{12 T_1}{539}+\frac{72 T_2}{539}-\frac{216 T_3}{539}+\frac{24 T_4}{49}+\frac{80 T_5}{539}-\frac{16 T_6}{539}-\frac{32 T_7}{49}\\
	&-\frac{208 T_8}{539}-\frac{8 T_9}{49}+\frac{16 T_{10}}{49} \,.
\eal
\subsection{Forms of the interacting part of the correlators}
\label{app:interactingparts}
In this section we present the forms of the interacting part of the correlators we have omitted in the main text, namely the raw result as it follows directly from our supergravity computation and its simplification after getting rid of big fractions. 
\paragraph{$\langle 2345 \rangle$}
Following the decomposition \eqref{eq:2345interacting} the $\mF$ is given by
\begin{eqnarray}
\mF(u,v) &= \left(\frac{5}{21} (u+1) v-\frac{650184197 v^2}{49545216}\right) \bar{D}_{2,4,5,5}+\frac{2}{7} u
   v \bar{D}_{2,4,4,4}-\frac{638387717 u v \bar{D}_{3,4,4,5}}{49545216} \nonumber\\
  &- \frac{10}{21} u v   \bar{D}_{3,4,5,4}-4 v \bar{D}_{1,3,4,4}+\frac{1}{14} v \bar{D}_{1,3,5,5}-\frac{1}{14} v
   \bar{D}_{1,4,4,5}-\frac{1}{21} v \bar{D}_{1,4,5,6} \nonumber\\
   &+\frac{656082437 v
   \bar{D}_{2,3,4,5}}{7077888}-\frac{2}{7} v \bar{D}_{2,3,5,4}-\frac{638387717 v
   \bar{D}_{2,3,5,6}}{49545216}-\frac{638387717 v \bar{D}_{2,4,4,6}}{49545216} \nonumber\\
   &-\frac{638387717 v
   \bar{D}_{3,3,4,6}}{49545216}-\frac{638387717 v \bar{D}_{3,3,5,5}}{49545216} \\
&=-\frac{26}{7}\, v\, \bar{D}_{1,3,4,4} +\frac{1}{14}\, v\,\bar{D}_{1,3,5,5}-\frac{5}{14}\, v\, \bar{D}_{1,4,4,5}-\frac{1}{21}\, v\,\bar{D}_{1,4,5,6}\nonumber\\
	&+\frac{5}{2}\, v\, \bar{D}_{2,3,4,5}+\frac{2}{7}\, u v\, \bar{D}_{2,4,4,4} + \frac{5}{21} (1 + u - v) v\, \bar{D}_{2,4,5,5}-\frac{10}{21}\, u v\, \bar{D}_{3,4,5,4}\,. \nonumber
\end{eqnarray}
The first form is the raw form of the correlator as we computed it from the effective supergravity Lagrangian. The second form is obtained by using $D$-function identities to get rid of the big fractions. 
\paragraph{$\langle 3456 \rangle$}
Following the decomposition in \eqref{eq:3456interacting} we can write the result in terms of three functions $\mathcal{F}_{1,2,3}$. From the supergravity computation we obtain the following complicated results for them:
\begin{equation}
\begin{aligned}
\mF_1 &= \frac{6}{7} u v^2 \bar{D}_{2,5,5,4}-\frac{5}{14} v^2 (-2 u+v-2) \bar{D}_{2,5,6,5}+\frac{62}{21}
   u v^2 \bar{D}_{3,5,5,5}-\frac{10}{7} u v^2 \bar{D}_{3,5,6,4} \\
   &-\frac{6273235427 u v^2
   \bar{D}_{4,5,5,6}}{1167851520}-\frac{55}{24} u v^2 \bar{D}_{4,5,6,5}+\left(\frac{3}{8} (u+1)
   v^2-\frac{6711179747 v^3}{1167851520}\right) \bar{D}_{3,5,6,6} \\
   &-12 v^2
   \bar{D}_{1,4,5,4}+\frac{3}{14} v^2 \bar{D}_{1,4,6,5}-\frac{3}{14} v^2
   \bar{D}_{1,5,5,5}-\frac{1}{7} v^2 \bar{D}_{1,5,6,6}-\frac{81}{7} v^2
   \bar{D}_{2,4,5,5} \\
   &-\frac{6}{7} v^2 \bar{D}_{2,4,6,4}-\frac{1}{3} v^2
   \bar{D}_{2,4,6,6}-\frac{13}{42} v^2 \bar{D}_{2,5,5,6}+\frac{1}{24} v^2
   \bar{D}_{2,5,6,7}+\frac{51114399029 v^2 \bar{D}_{3,4,5,6}}{908328960} \\
   &+\frac{11}{42} v^2
   \bar{D}_{3,4,6,5}-\frac{6273235427 v^2 \bar{D}_{3,4,6,7}}{1167851520}-\frac{6273235427 v^2
   \bar{D}_{3,5,5,7}}{1167851520}-\frac{6273235427 v^2
   \bar{D}_{4,4,5,7}}{1167851520} \\
   &-\frac{6273235427 v^2 \bar{D}_{4,4,6,6}}{1167851520}
\end{aligned},
\end{equation}
\begin{equation}
\begin{aligned}
\mF_2 &= \frac{1}{5} (2 u+2 v-1) \bar{D}_{2,5,4,7}+\left(\frac{53 u}{280}+\frac{53
   v}{280}-\frac{10853512121}{32699842560}\right) \bar{D}_{3,5,5,7}+\frac{3}{4} u
   \bar{D}_{2,5,3,6} \\
   &-\frac{3}{5} u \bar{D}_{3,5,3,7}+\frac{49}{20} u
   \bar{D}_{3,5,4,6}-\frac{83}{70} u \bar{D}_{4,5,4,7}-\frac{133254259 u
   \bar{D}_{4,5,5,6}}{934281216}-\frac{1}{4} v \bar{D}_{1,5,4,6} \\
   &-\frac{1}{10} v
   \bar{D}_{1,5,5,7}-\frac{7}{40} v \bar{D}_{2,5,5,6}+\frac{9}{140} v
   \bar{D}_{2,5,6,7}-\frac{133254259 v \bar{D}_{3,5,6,6}}{934281216}-\frac{27}{10}
   \bar{D}_{1,4,3,6} \\
   &+\frac{1}{4} \bar{D}_{1,4,4,7}-\frac{3}{4} \bar{D}_{2,4,3,7}-\frac{12}{7}
   \bar{D}_{2,4,4,6}-\frac{13}{40} \bar{D}_{2,4,5,7}-\frac{9}{20}
   \bar{D}_{3,4,4,7}+\frac{17159910329 \bar{D}_{3,4,5,6}}{3633315840} \\
   &-\frac{133254259
   \bar{D}_{3,4,6,7}}{934281216}-\frac{133254259 \bar{D}_{4,4,5,7}}{934281216}-\frac{133254259
   \bar{D}_{4,4,6,6}}{934281216}
\end{aligned}
\end{equation}
\begin{equation}
\begin{aligned}
\mF_3 &= \frac{4325188189 u^3 \bar{D}_{4,5,5,6}}{11678515200}+\frac{4325188189 u^2 v
   \bar{D}_{3,5,6,6}}{11678515200} \\
   &+\left(\frac{19181727883 u^2}{81749606400}+\frac{19}{140} u
   (v+1)\right) \bar{D}_{3,4,6,7}-\frac{12369260683 u^2
   \bar{D}_{3,4,5,6}}{9083289600} \\
   &+\frac{4325188189 u^2
   \bar{D}_{3,5,5,7}}{11678515200}+\frac{4325188189 u^2
   \bar{D}_{4,4,5,7}}{11678515200}+\frac{4325188189 u^2
   \bar{D}_{4,4,6,6}}{11678515200} \\
   &-\frac{3}{14} u v \bar{D}_{2,4,6,6}+\frac{3}{28} u v
   \bar{D}_{2,5,6,7}-\frac{3}{20} (u-2 (v+1)) \bar{D}_{2,3,6,7}-\frac{3}{5} u
   \bar{D}_{2,3,5,6}-\frac{47}{140} u \bar{D}_{2,4,5,7} \\
   &+\frac{39}{140} u
   \bar{D}_{3,3,5,7}+\frac{48}{35} u \bar{D}_{3,3,6,6}-\frac{117}{140} u
   \bar{D}_{4,3,6,7}-\frac{1}{10} v \bar{D}_{1,3,6,6}-\frac{1}{10} v
   \bar{D}_{1,4,6,7}-\bar{D}_{1,2,5,6} \\
   &+\frac{1}{10} \bar{D}_{1,3,5,7}-\frac{1}{5}
   \bar{D}_{2,2,5,7}+\frac{1}{5} \bar{D}_{2,2,6,6}-\frac{3}{10} \bar{D}_{3,2,6,7}. 
\end{aligned}
\end{equation}
They can be simplified using $D$-function identities to get rid of the huge rationals. They result in the following expressions:
\bal
	\mathcal{F}_1 &= -12\, v^2\, \bar{D}_{1,4,5,4}+\frac{3}{14} \,v^2 \,\bar{D}_{1,4,6,5}-\frac{3}{14}\, v^2\, \bar{D}_{1,5,5,5}-\frac{1}{7}\, v^2\, \bar{D}_{1,5,6,6}\\
	&+\frac{6}{7}\, u v^2\, \bar{D}_{2,5,5,4}-\frac{97}{6}\, v^2\, \bar{D}_{2,4,5,5}-\frac{6}{7}\, v^2\, \bar{D}_{2,4,6,4}+\frac{33}{14}\, v^2\, \bar{D}_{2,4,6,6}-\frac{13}{42}\, v^2\, \bar{D}_{2,5,5,6}\\
	& +\frac{1}{24}\, v^2\, \bar{D}_{2,5,6,7} + \frac{1}{21}\, v^2 \,(15 u-13 v+15) \bar{D}_{2,5,6,5}+\frac{3}{28}\, v\, (7 u+67 v+7) \bar{D}_{3,4,5,6}\\
	& +\frac{3}{8}\, v (u-v+1) \,\bar{D}_{3,4,6,7} -\frac{10}{7}\, u v^2 \,\bar{D}_{3,5,6,4}-\frac{55}{24} \,u v^2 \,\bar{D}_{4,5,6,5}-\frac{3}{8}\, v (u-v+1) \,\bar{D}_{3,5,5,7}\,,
\eal
\bal
	\mathcal{F}_2 &= -\frac{1}{4} \,v\, \bar{D}_{1,5,4,6}-\frac{1}{10}\, v \,\bar{D}_{1,5,5,7}-\frac{27}{10}\, \bar{D}_{1,4,3,6}+\frac{1}{4}\, \bar{D}_{1,4,4,7}\\
	& + \frac{1}{5}\, (2 u+2 v-1) \,\bar{D}_{2,5,4,7}+\frac{3}{4}\, u\, \bar{D}_{2,5,3,6}-\frac{7}{40} \,v\, \bar{D}_{2,5,5,6}+\frac{9}{140}\, v \,\bar{D}_{2,5,6,7}-\frac{3}{4}\, \bar{D}_{2,4,3,7}\\
	&-\frac{12}{7}\, \bar{D}_{2,4,4,6}-\frac{13}{40}\,\bar{D}_{2,4,5,7}-\frac{3}{5} \,u\, \bar{D}_{3,5,3,7}+\frac{49}{20} \,u \,\bar{D}_{3,5,4,6}-\frac{9}{20}\, \bar{D}_{3,4,4,7}+\frac{963}{280}\,\\
	&  +\frac{53}{280} \,(u+v-1)\, \bar{D}_{3,5,5,7} \bar{D}_{3,4,5,6} -\frac{83}{70} \,u \,\bar{D}_{4,5,4,7}\,,
\eal
\bal
	\mathcal{F}_3 &= -\frac{1}{10}\, v \,\bar{D}_{1,3,6,6}-\frac{1}{10}\, v\, \bar{D}_{1,4,6,7}-\bar{D}_{1,2,5,6}+\frac{1}{10}\, \bar{D}_{1,3,5,7}\\
	& -\frac{3}{14}\, u v \,\bar{D}_{2,4,6,6}+\frac{3}{28}\, u v\, \bar{D}_{2,5,6,7}-\frac{3}{5}\, u\, \bar{D}_{2,3,5,6}-\frac{47}{140} \,u \,\bar{D}_{2,4,5,7}-\frac{1}{5} \,\bar{D}_{2,2,5,7}+\frac{1}{5}\, \bar{D}_{2,2,6,6}\\
	& +\frac{69}{35} \,u^2 \,\bar{D}_{3,4,5,6}+\frac{39}{140}\, u\, \bar{D}_{3,3,5,7}+\frac{48}{35} \,u\,\bar{D}_{3,3,6,6}-\frac{117}{140}\, u\,\bar{D}_{4,3,6,7}-\frac{3}{10}\, \bar{D}_{3,2,6,7} \\
	&-\frac{3}{20}\, (u-2 (v+1)) \,\bar{D}_{2,3,6,7} -\frac{19}{140}\, u (u-v-1)\, \bar{D}_{3,4,6,7}\, .
\eal

\subsection{Free parts}
\label{app:freeparts}
In section \ref{sec:results} we have listed the interacting part of the correlators we have computed following the decomposition in section \ref{sec:generalities}. The complete correlator is the sum of a free and interacting part, thus to express the full correlator we need to compute the free part in the planar limit as well. As discussed in section \ref{sec:extendedness} the correct free part is computed from extended CPO's which we denoted by $\tilde{\mO}_k^{I_k}$. As a reference and empirical evidence for the reader concerning the effect of extendedness we list the planar limit of the free parts of all the non-trivial four-point functions $\langle k_1 k_2 k_3 k_4 \rangle$ with $k_i 
\leq 5$ in table \ref{tab:freeparts}. When a difference exists we list both the result for non-extended (upper part) and extended operators (lower part). Although in principle one can use combinatorics to compute the planar limit of the free part\footnote{A hint of this fact can be found in the regularity of the appearing numbers: after factoring an overall constant only simple integers remain.} we have computed them using a straightforward implementation in Mathematica of Wick contractions between scalars in $\mN=4$ SYM using the formulae in \cite{Kristjansen:2002bb}, with the exception of $\langle 5555 \rangle$ which we took from \cite{Dolan:2006ec}. The correlators which are not listed vanish identically. 
\renewcommand{\arraystretch}{1.8}
\begin{table}
\begin{center}
\begin{tabular}{ |c|p{12cm}| } 
 \hline
\textbf{ Correlator} & \textbf{Planar free part coefficients} $\tfrac{1}{N^2} \{c_{a_l}\}$ \\
 \hline
 $\langle 2222 \rangle$ & $\{N^2, 4, N^2, 4, 4, N^2\}$ \\ 
 \hline
  $\langle 2233 \rangle$ & $\{ 0, 12, 0, 6, 6, N^2 \}$ \\ 
 \hline
  $\langle 2244 \rangle$ & \parbox{10cm}{$\{ 16, 24, 16, 8, 8, N^2  \}$ \\ $\{ \phantom{0}0, 24, \phantom{0}0, 8, 8, N^2  \}$} \\ 
 \hline
  $\langle 2255 \rangle$ & \parbox{10cm}{$\{30, 40, 30, 10, 10, N^2  \}$ \\ $\{ \phantom{0}0, 40, \phantom{0}0, 10, 10, N^2  \}$} \\ 
 \hline  
  $\langle 2334 \rangle$ & \parbox{10cm}{ $\sqrt{2} \{12 , 12 , 0, 12 , 6 , 0  \}$ \\  $\sqrt{2} \{\phantom{0}0 , 12 , 0, 12 , 6 , 0  \}$} \\ 
 \hline
   $\langle 2345 \rangle$ & \parbox{10cm}{ $\sqrt{30} \{6 , 6 , 4 , 4 , 2 , 0  \}$ \\  $\sqrt{30} \{0 , 6 , 0 , 4 , 2 , 0  \}$}
   \\
 \hline
    $\langle 2444 \rangle$ & \parbox{10cm}{ $\sqrt{2} \{16, 16 , 16 , 16, 16 , 16 \}$ \\  $\sqrt{2} \{\phantom{0}0, 16 , \phantom{0}0 , 16, 16 , \phantom{0}0 \}$}
   \\
 \hline
  $\langle 2455 \rangle$ & \parbox{10cm}{ $\sqrt{2} \{30 , 30 , 30 , 20 , 20 , 20  \}$ \\  $\sqrt{2} \{\phantom{0}0 , 30 , \phantom{0}0 , 20 , 20 , \phantom{0}0  \}$}
   \\ 
 \hline
  $\langle 3333 \rangle$ & $\{ N^2, 9, 9, N^2, 9, 18, 9, 9, 9, N^2  \}$ \\ 
 \hline
  $\langle 3344 \rangle$ & $\{0, 24, 24, 0, 12, 24, 12, 12, 12, N^2  \}$ \\ 
 \hline
  $\langle 3355 \rangle$ & \parbox{10cm}{ $\{30,45,45,30,15,30,15,15,15,N^2 \}$ \\ $\{\phantom{0}0,45,45,\phantom{0}0,15,30,15,15,15,N^2 \}$} \\
 \hline
  $\langle 3456 \rangle$ & $\sqrt{10} \{ 0, 18, 18, 0, 12, 12 , 6, 12, 6, 0 \}$ \\ 
 \hline
  $\langle 3555 \rangle$ & \parbox{10cm}{ $\sqrt{15} \{10,10,10,10,10,10,10,10,10,10\}$ \\ $\sqrt{15} \{\phantom{1}0,10,10,\phantom{1}0,10,10,10,10,10,\phantom{1}0\}$} \\
 \hline
  $\langle 4444 \rangle$ & $\{N^2, 16, 16, 16, N^2, 16, 32, 32, 
16, 16, 32, 16, 16, 16, N^2 \}$ \\ 
 \hline
  $\langle 4455 \rangle$ & $\{0, 40, 40, 40, 0, 20, 40, 40, 20, 20, 40, 20, 20, 20, N^2\}$ \\ 
 \hline
  $\langle 5555 \rangle$ &  \scalebox{0.98}{$\{N^2, 25, 25, 25, 25, N^2, 25, 50, 50, 50, 25, 25, 50, 50, 25, 25, 50, 25, 25, 25, N^2\}$} \\ 
 \hline
\end{tabular}\caption{Free parts of all non-trivial four-point correlators with weights up to $5$ and that of $\langle 3456 \rangle$ in the leading order of the planar limit, split to the coefficients in its decomposition \eqref{eq:freeG} as a list $\tfrac{1}{N^2} \{c_{a_l}\}$ with $\tfrac{1}{N^2}$ factored out. If there is a difference between the correlator of non-extended operators and that of the extended ones we list both results with the non-extended correlator appearing first.}  
\label{tab:freeparts}
\end{center}
\end{table}
\newpage
\bibliography{bibliography}

\end{document}

%% file: preambleSS.tex
\pdfoutput=1

\usepackage{jheppub}
\usepackage{amsmath,amsfonts,mathrsfs,graphicx,bbm}
\usepackage{fixltx2e}
\usepackage{calligra}

\usepackage[font=small]{caption}

\usepackage{lmodern}

\usepackage{graphicx}
\usepackage{booktabs}
\usepackage{dsfont}

\usepackage{mathrsfs}
\usepackage{amsmath,amssymb}
\usepackage{mathtools}
\usepackage{bbm}
\usepackage{slashed}
\usepackage{braket}

\usepackage{hyperref}

\usepackage{xspace}

\usepackage{color}

\definecolor{grey}{rgb}{0.4,0.4,0.5}
\definecolor{darkgreen}{rgb}{0,0.5,0}
\definecolor{darkred}{rgb}{0.6,0.0,0}
\definecolor{lightbrown}{rgb}{1,0.9,0.8}
\definecolor{brown}{rgb}{0.6,0.3,0.3}
\definecolor{darkblue}{rgb}{0,0,0.8}
\definecolor{darkmagenta}{rgb}{0.5,0,0.5}

\def\de{\delta }

\newcommand{\sslash}{\mathbin{/\mkern-6mu/}}

\newcommand{\mO}{\mathcal{O}}
\newcommand{\mF}{\mathcal{F}}

\newcommand{\mI}{\mathcal{I}}
\newcommand{\mJ}{\mathcal{J}}
\newcommand{\mK}{\mathcal{K}}
\newcommand{\mN}{\mathcal{N}}

\numberwithin{equation}{section}

\makeatletter
 \let\old@startsection=\@startsection
 \let\oldl@section=\l@section
 \renewcommand{\@startsection}[6]{\old@startsection{#1}{#2}{#3}{#4}{#5}{#6\mathversion{bold}}}
 \renewcommand{\l@section}[2]{\oldl@section{\mathversion{bold}#1}{#2}}
\makeatother

\renewcommand{\leq}{\leqslant}
\renewcommand{\geq}{\geqslant}

\DeclareMathAlphabet{\mathcalligra}{T1}{calligra}{m}{n}

\def\XXint#1#2#3{{\setbox0=\hbox{$#1{#2#3}{\int}$}
    \vcenter{\hbox{$#2#3$}}\kern-.5\wd0}}

\newcommand{\AdS}{\text{AdS}}

\newcommand{\alg}[1]{\mathfrak{#1}}

\newcommand{\be}{\begin{equation}}
\newcommand{\ee}{\end{equation}}
\newcommand{\bea}{\begin{eqnarray}}
\newcommand{\eea}{\end{eqnarray}}
\newcommand{\bal}{\begin{equation}\begin{aligned}}
\newcommand{\eal}{\end{aligned}\end{equation}}

\newcommand{\bee}{\begin{enumerate}}
\newcommand{\eee}{\end{enumerate}}
\newcommand{\bei}{\begin{itemize}}
\newcommand{\eei}{\end{itemize}}

\newcommand{\ads}{${\rm  AdS}_5\times {\rm S}^5\ $}

\def\ov{\over}

\def\la{\label}

\def\a {\alpha}
\def\b {\beta}

\def\g {\gamma}
\def\de {\delta}
\def\De {\Delta}

\def\p{\phi}

\def\z{\zeta}

\def\r {\rho}
\def\m {\mu}
\def\G{\Gamma}

\def\pa {\partial}

\def\x'{\mathaccent 19 x}
\def\y'{\mathaccent 19 y}
\def\n'{\mathaccent 19 n}
\def\u'{\mathaccent 19 u}

\def\et'{\mathaccent 19 \eta}
\def\th'{\mathaccent 19 \theta}
\def\lam'{\mathaccent 19 \lambda}
\def\varet'{\mathaccent 19 \vartheta}
\def\rh'{\mathaccent 19 \rho}
\def\ph'{\mathaccent 19 \phi}
\def\xb'{\mathaccent 19 {\bar{x}}}

\def\l{{\lambda}}
\def\na{{\nabla}}

\def\N{${\cal N}=4$ }

\def\sl(2){\alg{sl}(2)}

\def\be{\begin{equation}}
\def\ee{\end{equation}}

\def\r {\rho}
\def\a {\alpha}
\def\b {\beta}

\def\pa {\partial}

\def\g {\gamma}

\def\p{\phi}
\def\la{\label}

\def\ov{\over}

\def\G{\Gamma}

\def\d1{{\dot{1}}}

\newcommand{\bem}{\left (\begin{matrix}}
\newcommand{\eem}{\end{matrix} \right )}

\usepackage{mciteplus}